\documentclass[twocolumn,aps,prb,groupedaddress]{revtex4-2}

\usepackage{chngcntr}

\usepackage{graphicx}
\usepackage{color}
\usepackage{amsmath}
\usepackage{enumitem}
\usepackage{amssymb}
\usepackage{hyperref}
\usepackage{cancel}
\usepackage{ulem}

\newcommand{\be}{\begin{equation}}
	\newcommand{\ee}{\end{equation}}

\newcommand{\bea}{\begin{eqnarray}}
	\newcommand{\eea}{\end{eqnarray}}

\newcommand{\p}{\partial}

\newcommand{\la}{\left\langle}
\newcommand{\ra}{\right\rangle}
\newcommand{\lb}{\left[}
\newcommand{\rb}{\right]}
\newcommand{\lp}{\left(}
\newcommand{\rp}{\right)}

\renewcommand{\vec}[1]{{\boldsymbol #1}}

\newcommand{\addKN}[1]{\textcolor{magenta}{#1}}
\newcommand{\addQ}[1]{\textcolor{red}{#1}}

\makeatother

\begin{document}
\title{
Nonlocal conductivity, continued fractions and current vortices in electron fluids}
\author{Khachatur G. Nazaryan and Leonid Levitov} 
\affiliation{Department of Physics, Massachusetts Institute of Technology, Cambridge, MA 02139}

\begin{abstract} 
Vortices in electron fluids are a key indicator of electron hydrodynamics. However, a comprehensive framework linking macroscopic vorticity measurements with microscopic interactions and scattering mechanisms has been lacking. We employ wavenumber-dependent conductivity $\sigma(k)$ incorporating realistic microscopic scattering processes, aiming to clarify the relationship between nonlocal response and vortices across ballistic and hydrodynamic phases. Vorticity is found to take similar values in both phases but feature very different sensitivity to momentum-relaxing scattering, with ballistic vortical flows being orders-of-magnitude more resilient than the hydrodynamic ones. This behavior can serve as a simple diagnostic of the microscopic origin of vorticity in electron fluids. 
%
\end{abstract}

\maketitle

Electron hydrodynamics is an emerging framework that describes interacting electron systems in a manner similar to conventional fluids, utilizing locally conserved quantities such as particle density and momentum to simplify the description at large length and time scales
\cite{Andreev2011, Levitov2016, 
Guerrero-Becerra2019,Hasdeo2021,Muller2009,Principi2016,Scaffidi2017, Narozhny2019,Alekseev2020,Toshio2020,Narozhny2021,Tomadin2014, Principi2016,Lucas2018,Qi2021,Cook2021,Valentinis2021a,Valentinis2021b, 
Energy_waves_2013,HGuo2017,Guo_thesis_2018,AShytov2018}. 
Hydrodynamics predicts 
remarkable effects like reduced dissipation and improved electrical conduction due to
carrier collisions\cite{HGuo2017,Kryhin2023b}, 
energy and heat propagating as waves \cite{Energy_waves_2013,Mazza2021,Zhao2023}, and vortices manifested through currents flowing against externally applied electric fields \cite{Levitov2016,Zeldov2022,Palm2024}, to name just a few. 
Vortices and their 
relationship with microscopic scattering processes and system geometry will be the focus of this study.




Spatial current patterns observed on macroscales, such as Stokes, Poiseuille, and vortical flows \cite{Levitov2016,Sulpizio2019,Ku2020,Braem2018,Vool2021,Zeldov2022,Palm2024}, 
encode information about carrier dynamics and interactions on microscales  \cite{Tomadin2014,Principi2016,Qi2021,Cook2021,Valentinis2021a,Valentinis2021b,Khoo2020,Khoo2021}. 
In particular, vortices, 
manifested through currents flowing against externally applied electric fields, attract interest as a telltale signature of 
electron viscosity \cite{Levitov2016,Bandurin2016,Pellegrino2016,Lucas2018}. 
Previous studies have often regarded vorticity as a distinct feature of the hydrodynamic phase. Here, we explore the conditions under which vortex patterns can arise in an electron system, focusing on laminar flows at low currents, which are relevant to ongoing experimental efforts 
\cite{Sulpizio2019,Ku2020,Braem2018,Vool2021}.

\begin{figure}[t]
\centering{}
\includegraphics[width=0.95\columnwidth]{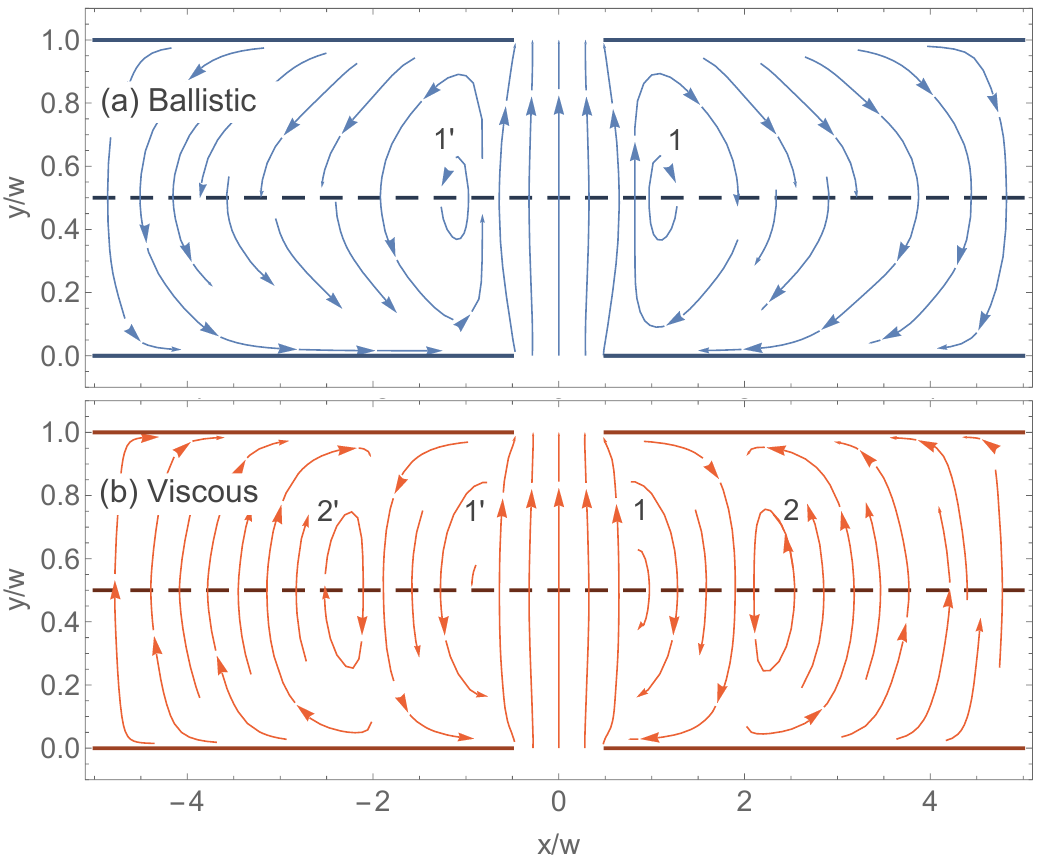} 
\caption{ 
%
Vortices in the ballistic (a) and viscous (b) regimes within a strip of width $w$  obtained using the nonlocal conductivity response, Eq.\eqref{eq:sigma_nonlocal}, with boundary conditions treated by the method outlined in Sec. \ref{sec:vortices1}. In both regimes, the $k$-dependent conductivity given in Eq.\eqref{sigma} was used, with the electron-electron collision rate  $\gamma=0.1\,v/w$ in (a) and $\gamma=100\,v/w$ in (b).
Currents flow into and out of the strip through a pair of slits on opposite sides, with the slit width set equal to the strip width. 
The two flows exhibit vortices of comparable intensity but different structure: one pair of vortices in the ballistic regime compared to two pairs in the viscous regime, centered at $x_{1,1'}\approx \pm w$ and $x_{2,2'}\approx \pm 2w$. 
} 
\label{fig:streams}
\end{figure}

As we will see, the requirements for vortices prove to be considerably less stringent than what previous work has suggested. Specifically, as illustrated in Fig.\ref{fig:streams}, robust vortical flows can arise not only in the viscous phase, but also in the ballistic phase, in which carrier-carrier scattering is inessential. Furthermore, vortical flows occurring outside the viscous phase are in general more resilient under the influence of momentum-relaxing scattering due to disorder and phonons. Different aspects of this behavior are discussed in Secs. \ref{sec:vortices1} and \ref{sec:general method}.


To investigate the hydrodynamic and ballistic phases, as well as the crossover between them, we require a method that treats these phases equally.
For this purpose, we employ a nonlocal conductivity framework described in Secs.\ref{sec:conductivity_and_vortices}, \ref{sec:quasineutrality} and \ref{sec:continued fractions}, which offers considerable flexibility by incorporating various factors such as different scattering mechanisms and multiple types of low-energy excitations. This approach is well-suited for exploring the ballistic-to-viscous crossover as it works equally well in both the ballistic and viscous phases. Unlike many studies of electron fluids that rely on minimal models such as the Navier-Stokes equation, which are effective only in the extreme hydrodynamic regime, this framework maintains validity across a wider range of conditions.

\section{Broad framework}
\label{sec:conductivity_and_vortices}

This section offers an overview of our approach, emphasizing the concepts while postponing detailed discussions to later sections.
We first introduce the nonlocal current-field response for systems without boundaries and discuss its properties in relation to current vortices.
The nonlocal conductivity relates the electric current at one point ($\vec r$) to the electric field at another point ($\vec r'$):
\be\label{eq:sigma_nonlocal}
j_\alpha(\vec r)=\int d^2x' \sigma_{\alpha\alpha'}(\vec r-\vec r')E_{\alpha'}(\vec r')
.
\ee 
The quantity of interest is the wavenumber-dependent and, in general, frequency-dependent conductivity 
\be\label{eq:sigma_alpha_beta}
\sigma_{\alpha\alpha'}(\vec k)=\int d^2 xe^{-i\vec k(\vec r-\vec r')} \sigma_{\alpha\alpha'}(\vec r-\vec r')
,
\ee 
which links Fourier harmonics of field and current. 
This response function, defined in the entire plane, in Sec.\ref{sec:continued fractions} will be linked, in a very general manner, to the rates of different microscopic scattering processes. Then,  in Secs. \ref{sec:vortices1} and \ref{sec:general method}, we discuss the methodology used to determine geometry-dependent flows in finite-size systems.

In the literature, nonlocal relations such as Eq.\eqref{eq:sigma_nonlocal} 
are found in the theory of nonlocal 
supercurrent response in clean superconductors \cite{Tinkham} and 
in the theory of anomalous skin effect \cite{Pitaevskii}. However, in these problems the nonlocal response describes 
supercurrents and normal currents in a narrow surface layer. To the contrary, in this study the relation in Eq.\eqref{eq:sigma_nonlocal} will be employed to describe currents in the bulk of the system. 
While nonlocal currents  are ubiquitous in 3D systems, 
the behavior in 3D bulk remains largely concealed by screening, and, to the best of our knowledge, has not been directly probed. To the contrary, in modern 2D materials electronic states are exposed and can be accessed by local probes, which makes these systems an appealing platform for studying the nonlocal current response \cite{Sulpizio2019,Ku2020,Braem2018,Vool2021,Zeldov2022,Palm2024}. 

Current vortices play a special role in this endeavor, offering distinct macroscopic signatures of nonlocal conductivity.  
Indeed, in conventional ohmic transport described by a local relationship between currents and fields, the flow is potential and vortex-free. Conversely, 
nonlocal conductivity, irrespective of the mechanisms involved, habitually leads to vortices. Studying the structure of vortices and their dependence on various experimental knobs can shed light on transport mechanisms leading to nonlocal conductivity. 
While the literature commonly explores currents and vortices in electron fluids within the context of viscous hydrodynamics (e.g., see Refs. \cite{Levitov2016, NoVort1, NoVort2}), it leaves unanswered questions regarding the persistence of vortices beyond the viscous phase.
Here, we explore the occurrence of vortices in the ballistic and viscous phases as well as at the crossover between these phases. 
As illustrated in Fig.\ref{fig:streams}, vortical flows are ubiquitous, emerging regardless of the specifics of the transport mechanism.
However, the structure of vortical flows (the numbers of vortices and their arrangment) 
reflects the $k$ dependence of conductivity, Eq.\eqref{eq:sigma_alpha_beta}. 

%
%
%
%



Next, we summarize the key aspects of the formalism:
\begin{enumerate}
\item Symmetry of the conductivity tensor, the role of screening and quasineutrality;
\item 
Deriving nonlocal conductivity from fermion kinetics: A continued fraction representation for $\sigma(k)$;
\item Boundary-value problem for nonlocal conductivity 
and implications for vortical flows. 
\end{enumerate}
Below we present an overview, for a detailed discussion the reader is referred to Secs.\ref{sec:quasineutrality}, \ref{sec:continued fractions}, \ref{sec:vortices1} and \ref{sec:general method}.

We consider symmetry of the conductivity tensor, assuming, for simplicity, a fully isotropic problem---namely, cylindrical symmetry of the bandstructure and interaction Hamiltonian under spatial rotations. 
Based on symmetry considerations, the conductivity 
describing the current-field response in the bulk of the system, can be expressed as the sum of its $\vec k$-transverse and $\vec k$-longitudinal components: 
\be\label{eq:sigma_perp_sigma_parallel}
\sigma_{\alpha\alpha'}(\vec k) =\sigma(k)(\delta_{\alpha\alpha'}-\hat{\vec k}_{\alpha}\hat{\vec k}_{\alpha'})
+ \sigma'(k) \hat{\vec k}_{\alpha}\hat{\vec k}_{\alpha'}
.
\ee 
This response function possesses 
full rotational symmetry. 
Assumed here for simplicity, the cylindrical symmetry case is relevant for many of the problems of current interest. 
The quantity in Eq.\eqref{eq:sigma_perp_sigma_parallel} can describe current response both in the AC regime at a finite frequency or in the DC regime at zero frequency, where the frequency dependence enters in $\sigma(k)$ and $\sigma'(k)$. The quantities $\sigma(k)$ and $\sigma'(k)$ can be found from the kinetic equation for quasiparticles, as described in Sec.\ref{sec:continued fractions}.

Further simplifications arise for the DC transport problem, which will be our focus below. 
In this case, the longitudinal part of the response vanishes, since longitudinal currents result in a rapid space charge buildup that generates fields by which these currents are quickly screened out. Indeed, a longitudinal electric field which is parallel to the wavevector, $\vec E\parallel\vec k$, is nothing but a gradient of a static cosine wave, $\phi(\vec r)\sim \cos(\vec k\vec r+\theta)$. Such potential, when applied to an electron system, is screened out by polarization in the electron gas and does not produce a DC current. Accordingly, in what follows we disregard the longitudinal response by setting $\sigma'(k) =0$.

It is instructive to consider more closely the physical origin of this screening effect. 
We first note that the conductivity $\sigma_{\alpha\alpha'}(\vec{k})$ given in Eq.\eqref{eq:sigma_perp_sigma_parallel}, in general, takes nonvanishing values for fields and currents
parallel and perpendicular to the wavevector $\vec{k}$. However, in considering the response of current to an external DC field, one must account for the
potential of a space charge that builds up due to the spatial nonuniformity
of current.
This is described by 
\begin{align}
 j_{\perp}(k)=\sigma(k)E_{\perp}(k),
 \quad j_{\|}(k)=\sigma'(k) (E_{\|}(k)-i\vec{k}\frac{\delta\mu(k) )}{e},
 \nonumber
\end{align}
where we added a term $\delta\mu(k)$ describing the
potential of a space charge buildup. 
In a steady state, the potential $\delta\mu(\vec r)$ 
induced by the longitudinal field component $E_{\|}$, determined by the continuity relation $\text{div}\vec{j}=0$, cancels $E_{\|}$. In other words, for time-independent fields and currents, the longitudinal currents vanish due to space charge buildup. As a result, DC currents are solely described by $\sigma(k)$ through a relation between the transverse components of $\vec{j}$ and $\vec{E}$, which involves
\begin{align}\label{eq:sigma(k)_general_tensor}
 & \sigma_{\alpha\alpha'}(\vec{k})=\sigma(k)\left(\delta_{\alpha\alpha'}-\hat{\vec{k}}_{\alpha}\hat{\vec{k}}_{\alpha'}\right)
 ,
\end{align}
that is by Eq.\eqref{eq:sigma_perp_sigma_parallel} with $\sigma'(k)=0$. 
Another key aspect of screening, arising due to ambient charges in the electron system rearranging in the presence of currents in order to maintain quasineutrality, will be discussed in Sec.\ref{sec:quasineutrality}. 

Next, we consider the conductivity $\vec k$ dependence. Microscopic analysis links the quantity $\sigma_{\alpha\alpha'}(\vec k)$ to the relaxation processes at the Fermi surface. Fermion kinetics that accounts for the dynamics of carrier distribution perturbed away from the Fermi sea equilibrium, 
yields a concise closed-form expression for 
conductivity $\sigma(k)$ in terms of the inverse lifetimes $\gamma_m$ 
for different angular harmonics of momentum distribution, organized in a continued fraction [see Sec.\ref{sec:continued fractions} and, also, Ref.\cite{Kryhin2023}]:
\begin{equation}\label{eq:sigma_transverse}
	\sigma(k) = \frac{D}{ \gamma_p+R(k)}
			,\quad
			    R(k)=\frac{z}{\gamma_2+\frac{z}{\gamma_3+\frac{z}{\gamma_4+\frac{z}{\gamma_5+...}}}}
		.
\end{equation}
Here $D=ne^2/m$ is the Drude spectral weight,  
and the wavenumber  dependence is encoded in the quantity $z=v_F^2 k^2/4$.

The dependence in Eq.\eqref{eq:sigma_transverse} has several remarkable properties. 
The sum of two distinct contributions $\gamma_p+R(k)$ in the denominator of $\sigma(k)$ 
describes momentum dissipation by momentum-relaxing and momentum-conserving processes, respectively. 
The first term is the rate $\gamma_p$ of 
scattering 
by disorder and phonons, the second term is a hydrodynamic dissipation rate $R(k)$
describing momentum loss due to carriers transporting it out of the region where the field induces current.

It is interesting to note the additive character of the contributions to resistivity $\sigma^{-1}(k)$
due to momentum-relaxing and momentum-conserving scattering. Namely, the quantities $\gamma_p$ and $R(k)$ obey a 'Matthiessen rule'. 
To the contrary, the contributions to resistivity from the perturbed Fermi surface angular harmonics are non-additive. Instead of adding up, the relaxation rates $\gamma_m$ for different angular harmonics combine through a continued fraction, as given in Eq.\eqref{eq:sigma_transverse}.

%
%
%

To clarify the role of the hydrodynamic momentum relaxation contribution $R(k)$, it is instructive to consider limiting cases of $\sigma(k)$. For spatially uniform fields and currents, $k=0$, the relation in Eq.\eqref{eq:sigma_transverse} reduces to the conventional local Drude conductivity. For a spatially nonuniform current flow, the quantity $R(k)$ describes a reduction in the conductivity at a nonzero $k$ arising from a nonlocal dissipation effect due to carriers moving away to the system boundary or contacts, where carriers' momenta can relax even if carrier-carrier scattering in the bulk of the system is momentum-conserving. 

This picture allows to understand the counter-intuitive effect of carrier collisions assisting conduction \cite{HGuo2017,Kryhin2023b}. Due to carrier collisions, each carrier spends more time within the system before it reaches the boundary or escapes out of the system. Higher collision rates therefore translate into smaller $R(k)$ values, suppressing dissipation and enhancing conductance. This explains, in very general terms, why the growth in the carrier-carrier collision rates with temperature assists conduction and reduces dissipation. 

Properties of $R(k)$ can be further illustrated by two examples. In the free-particle regime, $\gamma_p=\gamma_m=0$, evaluating continued fraction gives $R(k)=\sqrt{z}=kv/2$ (see Eq.\eqref{eq:Gamma(k)} and discussion beneath it). In this case, the response function $\sigma(k)$ describes ballistic transport in which the nonlocal conductivity is mediated by noninteracting particles freely moving through the system. Linear scaling $R(k)\sim k$ translates into conductivity $k$ dependence $\sigma(k)\sim 1/k$, which can be understood by considering ballistic transport through a slit or aperture of width $w\gg \lambda_F$. In this case the conductance is expected to scale linearly with the width $w$, which is precisely the scaling that follows from the relation $\sigma(k)\sim 1/k$. 

Another simple limiting case 
is when carrier collisions do occur but the wavenumbers $k$ are small. 
In this case, at leading order in $k^2$, disregarding contributions of $\gamma_m$ with $m>2$ yields $R(k)=\nu k^2$
where $\nu=v_F^2/4\gamma_2$. This gives the nonlocal conductivity which coincides with that obtained from viscous hydrodynamics, $\sigma(k)=D/(\gamma_p+\nu k^2)$, where $\nu$ is the kinematic viscosity. These two conductivity models will be further discussed below (see Eq.\eqref{eq:conductivities_a_b}) and employed to understand vortical flows. 

Lastly, we discuss the approach used below to tackle the boundary-value problem for nonlocal conductivity. 
In the literature on electron fluids, currents and vortices in systems of finite size are typically investigated using hydrodynamic models that employ partial differential equations (such as Navier-Stokes equations) with appropriately defined boundary conditions (see, for instance, Refs.\cite{Levitov2016,NoVort1,NoVort2,Kiselev2019,Asafov2022,Raichev2022}.
Treating nonlocal conductivity in a finite-size system represents a problem of a very different kind \cite{HGuo2017,Guo_thesis_2018,tomogrph,Kryhin2023,Qi2021}. In this context, it is natural to extend the problem to the entire 2D plane by considering currents and potentials equally both inside and outside the region of interest. In this approach, the physical boundary is replaced by a contour that nominally allows current flow but effectively acts as an impenetrable barrier due to its high resistivity. Mathematically, this concept is implemented by defining an auxiliary electric field localized on the boundary, which is treated as a free parameter and determined self-consistently to nullify currents at the impenetrable parts of the boundary. 


This approach enables the treatment of transport for general nonlocal conductivity, as given in Eq.\eqref{eq:sigma_transverse}, and accommodates arbitrary system geometries. This flexibility offers a significant advantage as it allows for the study of currents and vortices without the need to commit to a specific regime. Below, this approach will be employed to characterize vorticity and vortices across several different regimes, including the viscous and ballistic regimes, as well as the intermediate crossover regime.

We will find that vortices, rather than being unambiguously associated with viscous flows, are a generic property of systems with dispersive ($k$-dependent) conductivity that governs a nonlocal current-field response. 
To elucidate the properties of vortices in different transport regimes we
tune the system from the viscous regime, occurring at high electron-electron collision rates, to the ballistic free-electron regime. We find that 
vorticity does not disappear when the electron collision rate decreases. To the contrary, the vorticity experiences little change upon the viscous-to-ballistic crossover, taking similar values in the ballistic and viscous regimes. 

We illustrate this resilience of vorticity for a flow in a strip geometry pictured in Fig.\ref{fig:streams}, with carriers injected and drained through a pair of contacts positioned at the opposite sides of the strip. 
The flow was obtained by the method outlined in Sec.\ref{sec:vortices1} and discussed in subsequent sections. 
The flows pictured in (a) and (b) were found 
for the nonlocal conductivity model in Eq.\eqref{sigma} for parameter values corresponding to the ballistic and viscous regimes. 
Both flows feature vortices of comparable intensity but different structure: two pairs of vortices in the viscous regime at $x_{1,1'}\approx \pm w$ and $x_{2,2'}\approx \pm 2w$ vs. one pair in the ballistic regime. Interpolated current distributions 
shown in (a) and (b), while accurately representing the flow geometry, 
greatly exaggerate secondary vortices in (b). For a quantitatively accurate representation of the primary and secondary vortices, see Fig.\ref{fig:Supp_Middle}. 
Vorticity values,  quantified by the nonmonotonic part of 
the stream function related to backflow,  are overall quite similar in the two flows (as quantified in Fig.\ref{fig:NoDisorder}).

The robustness and generic character of vorticity in electron flows prompts a question of how the vortex patterns observed experimentally can be linked to the microscopic interactions and scattering mechanisms.  
Naively, judging from Fig.\ref{fig:streams} this may seem challenging. Indeed, despite somewhat different appearance 
in the viscous and ballistic phases, vortex patterns feature comparable vorticity values. 
However, while vorticity experiences little change upon the viscous-to-ballistic crossover, 
its response to momentum-relaxing collisions due to phonons or disorder is completely different in the two cases. 
Namely, vorticity is suppressed by momentum-relaxing scattering orders-of-magnitude more strongly in the viscous phase than in the ballistic phase. 
That is, a minuscule momentum-relaxing scattering is sufficient to suppress the vorticity of viscous flows, leaving vorticity of ballistic flows practically unaffected.
As discussed below, this 
behavior can 
serve as a diagnostic allowing to delineate between ballistic and viscous vortical flows. 

\section{
Linear response and quasineutrality}
\label{sec:quasineutrality} 
In this and the next section, we link the nonlocal conductivity $\sigma(k)$ to fermion kinetics, deriving the continued fraction representation as given in Eq.\ref{eq:sigma_transverse}. While the general form of this response function matches that obtained elsewhere \cite{Kryhin2023}, the treatment presented here addresses key questions such as the role of space charge and quasineutrality. 
As a general framework, we adopt the Boltzmann kinetic equation for carriers in the presence of an external electric field. Taking the free-particle Hamiltonian to be $H=\vec p^2/2m+U(\vec r)$ and accounting for relevant 
collision processes, the kinetic equation describes evolution of carrier distribution as
%
%
\begin{align}\label{eq:kinetic_eqn}
\frac{d f}{dt}+[f,H]=I(f)
.
\end{align}
Here $f(\vec p,\vec r,t)$ is electron momentum distribution evolving in space and time, and $[f,H]$ denotes the Poisson bracket 
\be\label{eq:Poisson_bracket}
\nabla_{\vec r} f\nabla_{\vec p}H-\nabla_{\vec r}H \nabla_{\vec p}f=\vec v\cdot\nabla_{\vec r} f+e\vec E\cdot\nabla_{\vec p}f
,
\ee 
where $\vec v$ and $e$ are the electron band velocity and charge. 
The collision term $I(f)$ will be discussed below. 

Nonlocal conductivity, describing linear response of currents to a weak external field $\vec E_{\rm ext} e^{i\vec k\vec r-i\omega t}$, can be found from perturbation in the carrier distribution $\delta f(\vec p,\vec r,t)$  induced by $\vec E_{\rm ext}$ throughout the system. After finding $\delta f(\vec p,\vec r,t)$ from the kinetic equation response, we will evaluate currents as
\be\label{eq:current_j(r)}
\vec j(\vec r,t)=\int \frac{d^2p}{(2\pi)^2} e\vec v(p)\delta f(\vec p,\vec r,t)
.
\ee
In general, the electric fields of interest are not purely potential, meaning that $\vec E_{\rm ext}$ has components both parallel and perpendicular to the modulation wavevector $\vec k$. To determine such response, we must consider the field $\vec E$ in Eq.\eqref{eq:Poisson_bracket} as the sum of an externally applied field $\vec E_{\rm ext}$ and an 'internal' field arising from the space charge (or polarization) induced by $\vec E_{\rm ext}$:
\be\label{eq:E+E'}
  \vec E(\vec r)=\vec E_{\rm ext}(\vec r)-\nabla_{\vec r}\int d^2x' U(\vec r-\vec r')\delta \rho(\vec r')
  ,
\ee
where $U(\vec r-\vec r')$ is the electron-electron $1/r$ interaction screened by gates and the dielectric environment. The perturbed density $\delta \rho$, which describes screening of the applied field by space charge buildup, must be determined selfconsistently from the kinetic equation response. 

Changes in the net carrier density due to currents flowing in the system are usually disregarded based on the `quasineutrality principle,' which assumes that any deviation from charge neutrality is screened out by ambient carriers. This concept is well established and widely used for local (ohmic) conductivity response in 3D metals. It is instructive to consider how it changes for the nonlocal response in 2D systems. To that end, we focus on the response of the carrier distribution to $\vec E_{\rm ext}(\vec r)$, including the effect of nonequilibrium space charge in the perturbed system as the second term in Eq.\eqref{eq:E+E'}, which hereafter we denote as $e\vec E_{U}(\vec r)$. 
We consider the kinetic equation linearized for a carrier distribution perturbed from the Fermi sea at equilibrium. Writing $f(\vec{p},\vec r,t)=f^{(0)}(\vec{p})+\delta f(\vec{p},\vec r,t)$, where $\delta f(\vec{p},\vec r,t)$ is first-order in $\vec E_{\rm ext}$, we have
\be\label{eq:kinetic_eqn_linearized}
 \left(\partial_{t}+\vec{v}\cdot\nabla_{\vec r}-I\right)\delta f 
 =-e(\vec{E}_{\rm ext}+\vec E_{U})\cdot\nabla_{\vec{p}}f_{\vec{p}}^{(0)}
 ,
\ee
where 
$I$ is a shorthand for the linearized collision operator 
and $f_{\vec{p}}^{(0)}$ is the equilibrium distribution. 

Using Eq.\eqref{eq:kinetic_eqn_linearized}, we can compare the solutions in the presence and in the absence of the interaction $U(\vec r-\vec r')$. It turns out that the interaction screens out the s-wave (angle-independent) part of momentum distribution, but does not impact other angular harmonics. As a result, since current obtained from Eq.\eqref{eq:current_j(r)} is insensitive to the angle independent part of momentum distribution, the current response to $\vec E_{\rm ext}$ obtained from Eq.\eqref{eq:kinetic_eqn_linearized} is identical to that obtained in the absence of interactions. 

To establish this behavior, we write momentum distribution as a sum of two terms
\be\label{eq:f=f(U=0)+df}
\delta f(\vec{p},\vec r,t)=\delta \tilde f(\vec{p},\vec r,t)+\frac{\p f_{\vec{p}}^{(0)}}{\p\epsilon} \hat U\delta\rho(\vec r,t)
,
\ee
where $\delta \tilde f$ defines an auxiliary `free-particle' carrier distribution that will be dealt with shortly, and the quantity $\hat U\delta\rho$ denotes a position-dependent potential given by the integral $\int d^2r' U(\vec r-\vec r')\delta \rho(\vec r')$. The last term in Eq.\eqref{eq:f=f(U=0)+df} can be formally interpreted as a change in the equilibrium carrier distribution due to a change in the 'chemical potential' by $\delta\mu(\vec r,t)=-\hat U\delta\rho(\vec r,t)$. 

To clarify the relation between the actual distribution $\delta f$, the fictitious free-particle distribution $\delta\tilde f$ and the quasineutrality picture, we integrate Eq.\eqref{eq:f=f(U=0)+df} over $d^2p$, finding a relation between densities
\be
\delta\rho(\vec r)=\delta \tilde\rho(\vec r)+\nu\int d^2r' U(\vec r-\vec r')\delta \rho(\vec r')
,
\ee
where $\nu=\int \frac{d^2p}{(2\pi)^2}(-\p f_{\vec{p}}^{(0)}/\p \epsilon)$ is the compressibility of the electron system. Introducing Fourier harmonics for densities and interaction, $\delta\rho(\vec k)=\int d^2r e^{-i\vec k\vec r}\delta\rho(\vec r)$, $\delta\tilde\rho(\vec k)=\int d^2r e^{-i\vec k\vec r}\delta\tilde\rho(\vec r)$, $U_{\vec k}=\int d^2r e^{-i\vec k\vec r}U(\vec r)$, we recover the standard Thomas-Fermi screening relation
\be\label{eq:Thomas-Fermi}
(1+\nu U_{\vec k})\delta\rho(\vec k)=\delta\tilde\rho(\vec k)
,
\ee
as expected for the `free-particle' density and the actual density affected by screening. For a long-range interaction, the relation in Eq.\eqref{eq:Thomas-Fermi} predicts strong screening. 

At the same time, perhaps surprisingly, the quantity $\delta \tilde f(\vec{p},\vec r,t)$ can be shown to obey a free-particle kinetic equation almost unaffected by screening. 
Indeed, by substituting into Eq.\eqref{eq:kinetic_eqn_linearized} the distribution $\delta f$ written as a sum given in Eq.\eqref{eq:f=f(U=0)+df}, we observe that, under $\vec v\cdot\nabla_{\vec r}$, the contribution $-\frac{\p f_{\vec{p}}^{(0)}}{\p\epsilon}\delta\mu$ cancels with the term $-e\vec E_{U}\nabla_{\vec{p}}f_{\vec{p}}^{(0)}$. We therefore obtain
\be
\partial_{t}\delta f 
+\left(\vec{v}\cdot\nabla_{\vec r}-I\right)\delta \tilde f 
=-e\vec{E}_{\rm ext}\cdot\nabla_{\vec{p}}f_{\vec{p}}^{(0)}
 ,
\ee
where we account for the fact that the quantity $\frac{\p f_{\vec{p}}^{(0)}}{\p\epsilon}$ is a zero mode of the linearized collision operator, replacing $I \delta f $ with $I \delta \tilde f $. For the DC transport problem, which is our focus in this study, the term $\partial_{t}\delta f $ drops out, indicating that the distribution $\delta\tilde f$ obeys a free-particle kinetic equation unaffected by interactions. A more complicated behavior is expected in the AC regime, since in this case oscillating space charge can excite collective plasma wave modes (to be discussed elsewhere). 
 
Properties of the linear response of the carrier distribution to $\vec E_{\rm ext}$ can be summarized by expressing the perturbed distribution in terms of modulations of the Fermi surface with different angular structure, provided by a sum of cylindrical harmonics
\be\label{eq:sum_delta f_m}
\delta f_{\vec{p}}(t,\vec r)=e^{i\vec{k}\vec r-i\omega t}\sum_{m}\delta f_{m}(p,t)e^{im\theta}
,
\ee 
where $\theta$ is the azimuthal angle on the Fermi surface and $p=|\vec p|$. The argument presented above proves that the $m\ne 0$ harmonics are unaffected by space charge buildup and screening, whereas the $m=0$ harmonic is reduced by the screening factor $1+\nu U_{\vec k}$. In the limit of a long-range interaction $U(\vec r-\vec r')$, the effect of screening is strong and the $m=0$ harmonic is nearly completely screened out. 


Mathematically speaking, the sharp difference between the $m=0$ and $m \ne 0$ harmonics arises because electron interactions are taken to be of a density-density form. As a result, only the density ($m=0$) harmonics are affected, whereas the $m \ne 0$ harmonics remain unchanged. If, instead, the interaction had some momentum dependence, as it does, for instance, in Fermi-liquid theory, the harmonics with nonzero $m$ would also be 
affected by screening.

It is worth noting that, while the predicted behavior aligns with the quasineutrality picture, there are some essential differences. Specifically, unlike the quasineutrality picture, the carrier distribution does not obey local equilibrium. Instead, as will become abundantly clear, the momentum distribution can be strongly angle-dependent. These effects become particularly pronounced in the ballistic transport regime discussed below. Therefore, the quasineutrality picture described here represents a significant extension of the usual quasineutrality framework. We thank Professor Emmanuel Rashba for clarifying to us the notion of space charge screening outside local equilibrium and its relation to quasineutrality, as summarized above.


%
%
%
%
%

\section{
conductivity $\sigma(k)$ and continued fractions}
\label{sec:continued fractions} 

Our next goal is to link the nonlocal conductivity to the properties of the linearized collision operator $I$. As we will see, it proves rewarding to focus on the normal modes and associated eigenvalues of this operator. The cylindrical symmetry of the problem allows us to work with the system of angular harmonics given in Eq.\eqref{eq
f_m}, each of which evolves in time as $e^{-\gamma_mt}$, where $\gamma_m$ are the eigenvalues of $I$. It turns out that the conductivity can be expressed directly in terms of the rates $\gamma_m$.
%

To set the stage for this discussion, we take a closer look at 
the collision operator $I(f)$ describing electron-electron scattering kinetics. 
In this case, 
\begin{align}\label{eq:collision_term}
&I_{\rm ee}(f_1)=
\sum_{21'2'}\lp w_{1'2'\to 12}-w_{12\to 1'2'}\rp
. 
\end{align} 
The quantities on the right-hand side of Eq.\eqref{eq:collision_term} describe the rate of change of the occupancy of a state $\vec p_1$, given as a sum of the gain and loss contributions resulting from the two-body scattering processes $12\to 1'2'$ and $1'2'\to 12$.  
The gain and loss contributions  
are related by the reciprocity symmetry $12\leftrightarrow 1'2'$. 
For these processes, Fermi's golden rule yields the scattering rates
\be\label{eq:Golden_Rule}
w_{1'2'\to 12}=\frac{2\pi}{\hbar}|V_{12,1'2'}|^2 
\delta_\epsilon \delta_{\vec p}(1-f_{1})(1-f_{2}) f_{1'}f_{2'},
\ee 
where the delta functions 
$\delta_\epsilon=\delta(\epsilon_1+\epsilon_2-\epsilon_{1'}-\epsilon_{2'})$, 
$\delta_{\vec p}=\delta^{(2)}(\vec p_1+\vec p_2-\vec p_{1'}-\vec p_{2'})
$ 
account for the energy and momentum conservation, the factors $1-f_1$ and $1-f_2$ account for fermion exclusion. Here $V_{12,1'2'}$ is the two-body interaction, 
properly antisymmetrized to account for fermion exchange and spin dependence (see, for instance, Ref.\cite{Ledwith2019}, Sec. 2). 
Interaction $V_{12,1'2'}$ depends on momentum transfer $k$ on the $k\sim k_F$ scale; 
this $k$ dependence is inessential and, for simplicity, can be ignored.  
The sum over momenta $2$, $1'$, $2'$ in Eq.\eqref{eq:collision_term} represents a six-dimensional integral over $\vec p_2$, $\vec p_{1'}$ and $\vec p_{2'}$, given by $(2\pi)^{-6}\int d^2p_2d^2p_{1'}d^2p_{2'}$ .

To understand the properties of the collision term in Eq.\eqref{eq:kinetic_eqn},
we briefly consider a spatially uniform problem, setting $[f,H]=0$. 
Using the standard ansatz 
\begin{align}
\delta f(\vec p)=-\frac{\p f^{(0)}_{\vec p}}{\p\epsilon}\eta(\vec p)
\end{align} 
yields a linear integro-differential equation for $\eta_1(\vec p,t)$: $f_{0}(1-f_{0})\frac{d \eta_1}{dt}=I_{\rm ee}\eta$ with the linearized collision operator 
%
\be\label{eq:I_ee}
I_{\rm ee}\eta=\sum_{21'2'} 
\lambda 
F_{121'2'} 
\delta_\epsilon \delta_{\vec p}
\lp \eta_{1'}+\eta_{2'}-\eta_{1}-\eta_{2}\rp
.
\ee
Here $\lambda$ 
denotes the interaction matrix element $\frac{2\pi}{\hbar}|V_{12,1'2'}|^2$,  
the quantity $F_{121'2'}$ is 
a product of the equilibrium 
Fermi functions
$f^0_{1}f^0_{2}(1-f^0_{1'})(1-f^0_{2'})$. 

Different excitations are described by eigenfunctions of the operator $I_{\rm ee}$, 
with the eigenvalues giving the decay rates equal to inverse lifetimes. Because of the cylindrical symmetry of the problem, the eigenfunctions are products of angular harmonics on the Fermi surface and functions of the radial energy variable $x=\beta(\epsilon-\mu)$:
\be
\eta(\vec p,t)= \sum_m e^{-\gamma_m t} e^{i m\theta}\chi_m(x)
,
\ee
where $\gamma_m$ and $\chi_m(x)$ are solutions of spectral problems 
\begin{align}
-\gamma_m 
f_{0}(1-f_{0}) e^{i m\theta}\chi_m(x)=I_{\rm ee}e^{i m\theta}\chi_m(x)
,
\end{align} 
one per each angular momentum channel.
The quantities $\gamma_{m}$ define inverse lifetimes of excitations in the Fermi gas representing modulations of the Fermi surface with different angular structure. 
The values $\gamma_m$, which give the spectrum of relaxation times of the system, have been studied some years ago in 3D Fermi liquids \cite{BaymPethick} and, recently, in 2D Fermi liquids \cite{Ledwith2019,tomogrph,Kryhin2022,Kryhin2023b,DasSarma2022}. These studies support the picture that 
in each angular momentum channel the modes with the smallest values $\gamma_m$ dominate the dynamics in each channel. At small temperatures, $\gamma_m$ become small, scaling as $T^2$ in 3D systems and as $T^4$ in 2D systems. Focusing on such longest-lived modes simplifies the picture and allows to arrive at a simple and general results for conductivity. 

Notably, a similar approach can be developed to describe other types of collisions, such as momentum-relaxing scattering by disorder and phonons. Due to the cylindrical symmetry of the problem, different eigenfunctions of the linearized collision operator can always be chosen as angular harmonics $\delta f_{m}(p)e^{im\theta}$, where
\be 
I \delta f_{m}(p)e^{im\theta}=-\gamma_m \delta f_{m}(p)e^{im\theta}
\ee 
(as above, we suppress the time dependence to emphasize that the operator $I$, which is nonlocal in $p$, is local in time and time-independent). This allows to incorporate these processes on equal footing with momentum-conserving electron-electron scattering
into a single framework. In particular, $\gamma_1=\gamma_{p}$ describes momentum relaxation due to disorder of phonon scattering, $\gamma_2$ is usually dominated by electron-electron collisions, $\gamma_0=0$ due to particle number conservation, and so on.



The key observation that allows progress towards a simple physical picture is that by choosing the basis $\delta f_{m}(p)e^{im\theta}$, the problem can be transformed into a tridiagonal matrix form. In this representation, which can be viewed as a one-dimensional tight-binding problem defined on a chain of sites labeled by different $m$, a closed-form solution for conductivity $\sigma(k)$ can be given in terms of continued fractions. This representation is derived from the analysis of the angular dependence of $\delta f_{\vec{p}}$ on the  Fermi surface, and the couplings between harmonics $m$ and $m\pm 1$  originating from the term $\vec v\cdot\nabla_{\vec r}$ in the kinetic equation. 

As a first step, we 
express the electric field term 
through carrier velocity as 
$\vec{E}\nabla_{\vec{p}}f_{\vec{p}}^{(0)}=\vec{E}\vec{v}\frac{\partial f_{\vec{p}}^{(0)}}{\partial\epsilon}$. 
This representation indicates that 
the term $e \vec{v} \vec{E}$, when rewritten in the angular harmonics basis, has nonzero matrix elements only between harmonics $m$ and $m\pm 1$. This is made apparent by the identities 
\begin{align}
e \vec{v} \vec{E}
&=\frac{ev}{2}\left(E_{x}+i E_{y}\right) e^{-i \theta}+\frac{ev}{2}\left(E_{x}-i E_{y}\right) e^{i \theta}
\nonumber \\
&=\mathcal{E} e^{-i \theta}+\bar{\mathcal{E}}e^{i \theta}
,
\end{align}
where we introduced notation 
$\mathcal{E},\bar{\mathcal{E}}=ev\left( E_x\pm i E_y\right)/2$. 
Likewise, the angular dependence of the streaming term $\vec{v} \vec{k}$ indicates that, when transformed to the angular harmonics basis, it also has nonzero matrix elements only between $m$ and $m\pm1$. Indeed,
\begin{align}
\vec{v} \vec{k} 
=\zeta e^{-i \theta}+\bar{\zeta} e^{i \theta} 
,\quad
\zeta,\bar\zeta=v(k_x\pm ik_y)/2
.
\end{align}
Accordingly, the Boltzmann equation becomes a system of coupled linear equations:
\begin{align}\label{eq:coupled_eqs}
& (\gamma_m-i\omega) \delta f_m  + \zeta\delta f_{m+1}+\bar{\zeta} \delta f_{m-1} =s_m
,
\\ \nonumber  & s_m = \frac{\partial f_{\vec p}^{(0)}}{\partial \epsilon} \left(\mathcal{E} \delta_{m,-1} + \bar{\mathcal{E}}\delta_{m,1}\right) 
,
\end{align}
where 
from now on we suppress, for conciseness, the dependence on $p$ in $\delta f_m$. This problem describes a response of variables $\delta f_m$ to the `source term' $s_m$ describing the electric field. 

While our main focus is the DC transport problem, it is instructive to consider a more general problem describing the response at a finite frequency. With this in mind, in Eqs.\eqref{eq:coupled_eqs}, we replace the rates $\gamma_m$ with $\gamma_m-i\omega$. This allows us to determine the nonlocal conductivity while maintaining full frequency dependence and then specialize to the DC case, $\omega=0$, at the end.

To solve these equations, we first 
consider the source term $s_m$ with $m=1$, adding the contribution of the $m=-1$ source term later. By introducing notation $\alpha_m=i \delta f_{m+1}/\delta f_m$, we can write 
equations with $m>1$ as 
\begin{align} 
  \tilde \gamma_m+\zeta\alpha_m-\frac{\bar{\zeta}}{\alpha_{m-1}}=0
  ,\quad \tilde \gamma_m=\gamma_m-i\omega
.
  \end{align}  
  These equations can be rewritten as recursion relations $\alpha_{m-1}=\frac{\bar{\zeta}}{\tilde\gamma_m+\zeta\alpha_m}$ and solved by iterations over $m+1$, $m+2$,$\dots$, yielding a continued fraction
  \begin{align}
  \alpha_{m-1}=\frac{\bar{\zeta}}{\tilde\gamma_m+\frac{\left|\zeta\right|^2}{\tilde\gamma_{m+1}+\frac{\left|\zeta\right|^2}{\tilde\gamma_{m+2}+\dots}}}
  .
  \end{align}
 Likewise, for $-\infty<m<1$ we define $\alpha'_m=i \delta f_{m-1} / \delta f_{m}$ and obtain 
  \begin{align} 
  \alpha'_{m+1}=\frac{\zeta}{\tilde\gamma_{m}+\frac{\left|\zeta\right|^{2}}{\tilde\gamma_{m-1}+\frac{\left|\zeta\right|^{2}}{\tilde\gamma_{m-2}+\ldots}}}
  .
  \end{align}
  Now, the harmonic $\delta f_{1}$ describing the current density can be found from the $m=1$ equation 
  \be\label{eq:df1+df2+df0} 
  \tilde\gamma_{1} \delta f_{1}+i \zeta \delta f_{2}+i \bar{\zeta} \delta f_{0}=\frac{\partial f_{\vec p}^{(0)}}{\partial \epsilon}{\mathcal{E}}.
  \ee 
  From now on, we impose the quasineutrality condition by suppressing the density harmonic $\delta f_{0}$. This quantity is small due to the space charge screening effects discussed in Sec.\ref{sec:quasineutrality} and, in what follows, will be set to zero.
  
After eliminating $\delta f_0$, the result in Eq.\eqref{eq:df1+df2+df0} takes the form 
$\delta f_{1}\left(\tilde\gamma_{1}+\zeta \alpha_{1} 
  \right)=\frac{\partial f_{\vec p}^{(0)}}{\partial \epsilon}\mathcal{E}$. Then, substituting the continued fraction for $\alpha_{1}$ 
  yields
\begin{align}\label{eq:f1_contfrac}
\delta f_{1}&=\frac{\partial f_{\vec p}^{(0)}}{\partial \epsilon}\frac{\bar{\mathcal{E}}}{\tilde \gamma_1+\frac{|\zeta|^{2}}{\tilde \gamma_{2}+\frac{|\zeta|^{2}}{\tilde \gamma_{3}+\frac{|\zeta|^{2}}{\tilde \gamma_{4}+\ldots}}}
}
. 
\end{align}
A similar result can be obtained for $\delta f_{-1}$ in terms of $\gamma_{-m}$ and $\mathcal{E}$. Current components can now be calculated from Eq.\eqref{eq:current_j(r)}. 
Introducing, for conciseness, the complex-valued quantities $j_{x}-ij_{y}$ and $v_x(p)-iv_y(p)=v(p)e^{-i\theta}$, gives the relation
\be
j_{x}-ij_{y}=
\la e(v_x(p)-iv_y(p))\lp \delta f_{1}e^{i\theta}+\delta f_{-1} e^{-i\theta}\rp\ra
\ee
where $\la...\ra$ denotes $\int \frac{d^2p}{(2\pi)^2}...$.
Taking the integral over $d^2p$ and noting that the contribution of $\delta f_{-1} $ vanishes upon integration over $\theta$, we obtain conductivity 
expressed as a continued fraction:
\begin{equation}\label{eq:sigma(k,w)}
	\sigma(k,\omega) = \frac{D}{ \tilde\gamma_1+\frac{z}{\tilde\gamma_2+\frac{z}{\tilde\gamma_3+\frac{z}{\tilde\gamma_4+...}}}}
			,\quad 
		z=v_F^2 k^2/4
		,
\end{equation}
where $D=ne^2/m$ is the Drude spectral weight. This result is identical to that in Eq.\eqref{eq:sigma_transverse}, with the carrier density $n=g k_F^2/4 \pi \hbar^2$, where $g$ is the spin/valley degeneracy and $k_F$ is Fermi momentum.  As discussed above, the term $\tilde\gamma_1$ describes the ordinary ohmic contribution, whereas the second term (denoted $R(k)$ in Eq.\eqref{eq:sigma_transverse}) describes a hydrodynamic contribution to dissipation due to the momentum transported by carriers lost outside the system.

%
%
%

Here we focus on a simple model for DC conductivity in which all rates $\gamma_{m\ge 2}$ are equal, $\gamma_2=\gamma_3=\gamma_4=
...=\gamma$ and we set $\omega=0$. 
In this case 
the quantity $R(k)$ can be obtained in a closed form. 
From the recursion relation $R(k)=z/(\gamma+R(k))$ we find
\begin{align} 
 R(k)=\frac{\sqrt{\gamma^{2}+k^{2}v^{2}}-\gamma}{2}
 ,
 \label{eq:Gamma(k)}
\end{align}
which gives a $k$-dependent DC conductivity 
\begin{align} 
 \sigma(k)=\frac{D}{\gamma_{p}+\frac12(\sqrt{v^2 k^{2}+\gamma^2}-\gamma ) }
 \label{sigma}
.
\end{align}
Here, for consistency with notation used in Eq.\eqref{eq:sigma_transverse} and throughout the text, we replaced $\gamma_1$ with the momentum relaxation rate $\gamma_{p}$.
This model, which describes the viscous and ballistic regimes, as well as the crossover between these regimes at the lengthscales such that $kv\sim\gamma$, will be used below to obtain vortical flows and analyze their stability in the presence of disorder scattering. 

This result for conductivity does not depend on the angle between electric field and the wave vector. As discussed above, accounting for the fact that the electric field component parallel to the wave vector $\vec k$ is screened out by space charge buildup and does not produce a DC current, yields an additional tensor structure $\delta_{\alpha \alpha^\prime} - \hat{\vec k}_\alpha \hat{\vec k}_{\alpha^\prime}$ and leads to a conductivity tensor given in Eq.\eqref{eq:sigma(k)_general_tensor}. The significance of this tensor structure depends on the system and current flow geometry. In a long strip with Poiseuille-type current flow directed along the strip, this tensor structure would be irrelevant, 
since in this case the field and current are directed perpendicular to the characteristic wavevector (see \cite{tomogrph,Kryhin2023}). However, the tensor structure $\delta_{\alpha \alpha^\prime} - \hat{\vec k}_\alpha \hat{\vec k}_{\alpha^\prime}$ will be key in the analysis of vortical flows studied below.  It also proves to be crucial in other geometries of interest, such as transport through a constriction \cite{HGuo2017,Guo_thesis_2018,Qi2021} or in a Corbino disk geometry.

Before concluding the discussion on the general properties of $\sigma(k)$, it is interesting to note that, as mentioned in the discussion following Eq.\eqref{eq:coupled_eqs}, this analysis can be adapted with minimal changes to describe current response at a finite frequency.
In that, the general symmetry-enforced decomposition of the conductivity tensor into the longitudinal and transverse components, given in Eq.\eqref{eq:sigma_perp_sigma_parallel}, remains unchanged, however both $\sigma(k)$ and $\sigma'(k)$ become functions of frequency. Frequency dependence of $\sigma(k)$ can be obtained as discussed above. The end result is Eq.\eqref{eq:sigma_transverse} in which the rates $\gamma_m$ are replaced with $\gamma_m-i\omega$. In this form, it provides an extension of the Drude-Lorentz model to the hydrodynamic domain. The longitudinal conductivity $\sigma'(k)$ changes in a different way, acquiring a plasmonic pole. This leads to a number of effects of interest for the AC response that will be discussed elsewhere. 

\section{Conductivity $\sigma(k)$ and vortical flows} 
\label{sec:vortices1}
While nonlocal conductivity is a key prerequisite for vortices, the geometry specifics are of course no less important. E.g. the strip geometry supports vortical flows \cite{Levitov2016,Semenyakin2018}, whereas the extensively studied open halfplane geometry supports vortex-free flows  \cite{NoVort1,NoVort2}. 
It is therefore of interest to identify a simple framework in which the general properties of vortical flows can be understood.

Here we develop an approach through which the nonlocal conductivity 
introduced above for a flow in an infinite plane can be used to find the flow in a system with boundaries. 
Namely, rather than specializing to specific system geometries, we continue to 
work in an infinite plane and mimic 
boundary conditions by adding to an externally applied field a fictitious field $\vec E^{({\rm fic})}$ concentrated on a line, or a system of lines, representing system boundary and chosen such that currents vanish at this `boundary'. For instance, to tackle the boundary value problem for the strip geometry pictured in Fig.\ref{fig:streams}, we solve the nonlocal conductivity problem in an infinite plane assigning high resistivity to the impenetrable parts of the strip boundary in Fig.\ref{fig:streams}. 
Field and current distributions are then determined from a selfconsistent solution to the transport problem in an infinite plane augmented by a fictitious boundary field $\vec E^{({\rm fic})}$. Namely, we employ Eq.\eqref{eq:sigma_nonlocal} with $\vec E$ corrected by a fictitious field $\vec{E}^{({\rm fic})}$ defined at the boundary and proportional to local current: 
\be
E_{\alpha'}(\vec r)=E_{\alpha'}^{({\rm ext})}(\vec r)-E_{\alpha'}^{({\rm fic})}(\vec r), 
\quad
E_{\alpha'}^{({\rm fic})}\left(\vec r\right)=\lambda j_{\alpha'}(\vec r),
\ee 
with $\lambda$ a `boundary resistivity' parameter. 
This yields a self-consistent integral equation for currents, 
\be\label{eq:j=sigma(E_ext-E_fic)}
j_{\alpha}(\vec r)=j_{\alpha}^{({\rm ext})}(\vec r)-\lambda \int_\mathcal{B} d\ell' \sigma_{\alpha\alpha'}(\vec r-\vec r') j_{\alpha'}(\vec r')
,
\ee
with the line integral taken along the system boundary $\mathcal{B}$.
We solve this equation by a method applicable for any
$\lambda$, setting it at the end
to a large enough value to mimic an impenetrable boundary 
(here, a value $\lambda=10^{5} v/Dw$ was used). 
The externally applied current is taken to be uniform, $\vec j^{({\rm ext})}(\vec r)=\vec j_{0}$.

The condition of zero currents at the boundary, enforced by a large $\lambda$, mimics the no-slip boundary condition in fluid mechanics. In this framework, more general boundary conditions can be introduced by using a symmetric $2\times2$ tensor quantity $\lambda_{\alpha\alpha'}$ with major axes normal and tangential to the boundary. Physically, the tensor $\lambda_{\alpha\alpha'}$ describes anisotropic resistivity at the boundary line. A large eigenvalue $(\lambda_\perp$ will enforce zero current normal to the boundary, whereas the tangential current component can be varied by tuning the eigenvalue $\lambda_\parallel$. Here, however, we focus on the isotropic case, where $\lambda_\perp$ and $\lambda_\parallel$ are equal and large. Other methods to tackle boundary conditions in electron hydrodynamics have been discussed in Refs.~\cite{Kiselev2019,Asafov2022,Raichev2022}.

The integral equation given in Eq.\ref{eq:j=sigma(E_ext-E_fic)}, which links quantities in the system interior and at the boundary, can be tackled in two steps. We first restrict $\vec r$ to system boundary. This yields a 1D integral equation for the boundary currents, a closed-form problem that can be solved directly. Next, the relation in Eq.\ref{eq:j=sigma(E_ext-E_fic)} between currents in the system interior and at the boundary can be used to predict currents 
in the entire 2D plane.


This approach has been established in previous literature. Recent examples are Refs.\cite{HGuo2017,Guo_thesis_2018,tomogrph,Kryhin2023,Qi2021}, and the first usage goes back as far as Ref.\cite{Levinson1977}. 
Furthermore, Ref.\cite{HGuo2017} provided a detailed comparison to other methods  
by using analytic solutions for the limiting cases.  
We note that, while 
the analysis below focuses on a simple strip geometry, the method outlined above is applicable to systems with arbitrary curved boundaries. 

On a more general note, this method  provides a simple, general and versatile framework for tackling boundary-value problems for nonlocal conductivity, not limited to the transport models discussed in Refs.\cite{HGuo2017,tomogrph}. While being heuristic, it is physically well motivated and intuitive. Though perhaps at present time this method is lacking rigorous justification, it gives results that agree with other approaches and, 
notably, with analytic solutions when these are available. An added value of this method is that it is versatile, flexible and insensitive to microscopic details, as will be made clear in Sec.\ref{sec:general method}. 


Using this approach and focusing on the $k$-dependent conductivity given in Eq.\eqref{sigma}, we will establish  
nonlocal conductivity as the key property responsible for the formation of vortices. 
As we will see, vortical flows 
emerging in a very general manner regardless of the specifics of the transport mechanism. In comparison, for a $k$-independent conductivity, which defines ohmic transport with a local current-field relation, 
the flow is potential and vortex-free.

One interesting aspect of vortical flows that will be elucidated by this study is that different transport mechanisms (viscous, ballistic, or else) result in different sensitivity of vortical flows to carrier momentum relaxation due to disorder or phonons. Properties of vorticity such as its robustness under the influence of disorder can be inferred directly from the conductivity $k$ dependence at small $k$. This is well illustrated by the $k$-dependent conductivity $\sigma(k)$, Eq.\eqref{sigma}, which defines a scale-dependent linear response that features different character at different lengthscales. 

The scale dependence and its relation to vorticity suppression by disorder is made more clear by writing $\sigma(k)=D/(\gamma_p+R(k))$, where $R(k)$ is given in Eq.\eqref{eq:Gamma(k)}. 
At small $k$ such that $R(k)<\gamma_{p}$ it describes ohmic dissipation. 
At large $k$ such that $R(k)>\gamma_{p}$ it describes dissipation due to particles transporting momentum out of the system. 
The large-$k$ behavior can be either ballistic or viscous-like, depending on the ratio of $\gamma$ and $kv$. 
Namely, as discussed above, for $\gamma \ll vk$ we have $R(k)=|\vec k|v/2$, 
whereas for $\gamma \gg vk$ we have $R(k)=k^2v^2/4\gamma$.
This gives two models of dispersive conductivity : 
\be\label{eq:conductivities_a_b}
{\rm a)}\ \sigma_{\rm ball}( k)=\frac{D}{\gamma_{p}+\frac{v|\vec k|}2}
;\quad
{\rm b)}\ \sigma_{\rm visc}( k)=\frac{D}{\gamma_{p}+\nu  k^2}
\ee
with $D=ne^2/m$ the Drude weight and $\nu =v^2/4\gamma$ the kinematic viscosity. Here $\gamma_{p}$ is the momentum relaxation rate due to disorder, $\gamma$ is the electron-electron collision rate that governs viscosity, and in the viscous case the long wavelength limit $kv\ll\gamma$ is assumed. 
The quantity in the denominators is the disorder scattering rate $\gamma_{p}$ corrected by a $k$-dependent contribution describing momentum relaxation due to momentum spreading over the lengthscales $\ell \sim 1/k$. 
The $k$ values for which this contribution becomes smaller than $\gamma_{p}$ define the lengthscales at which the conductivity becomes effectively local, yielding a current flow that is potential and vortex-free. 

For transport in a system of size $w$, the wavenumbers describing the flow that carries momentum out of the system are on the order $k\sim 1/w$. 
Compared to Eq.\eqref{eq:conductivities_a_b}, this predicts threshold values for disorder scattering above which the $k$ dependence of conductivity is suppressed:
\be
{\rm a)}\ \gamma_{p}'\approx v/w
;\quad
{\rm b)}\ \gamma_{p}''\approx \nu/w^2
,
\ee
respectively for the ballistic and viscous regimes. 
Condition a) states that vorticity is suppressed when the disorder mean free path is smaller than the system size. Condition b) states that the momentum relaxation time is shorter than the time momentum diffuses across viscous fluid in a system of size $w$, which is a considerably more stringent condition than a). 
These two threshold values are related as
\be
\gamma_{p}''/\gamma_{p}'\approx \ell_{\rm ee}/w
,
\ee
where $\ell_{\rm ee}=v/\gamma$ is the el-el collision mean free path. We see that in a hydrodynamic regime, $\ell_{\rm ee}\ll w$, the sensitivity of vortices to momentum-relaxing collisions is orders of magnitude stronger than in the ballistic regime. 

Physically, the reason for this difference is that in the viscous phase particles move along zigzagging paths that are considerably longer than the straight paths in the ballistic regime. Indeed, for vorticity to be insensitive to disorder the travel time over a typical path must be shorter than momentum relaxation time. We therefore expect ballistic vortices to be considerably more resilient under momentum-relaxing scattering than the viscous vortices. 
The general character of these conclusions, valid for a general dispersion $\sigma(k)$, will be confirmed by the numerical results presented in Sec.\ref{sec:general method}. 

\section{The influence of disorder} 
\label{sec:general method}

\begin{figure}[!t]
\centering{}\includegraphics[width=0.925\columnwidth]{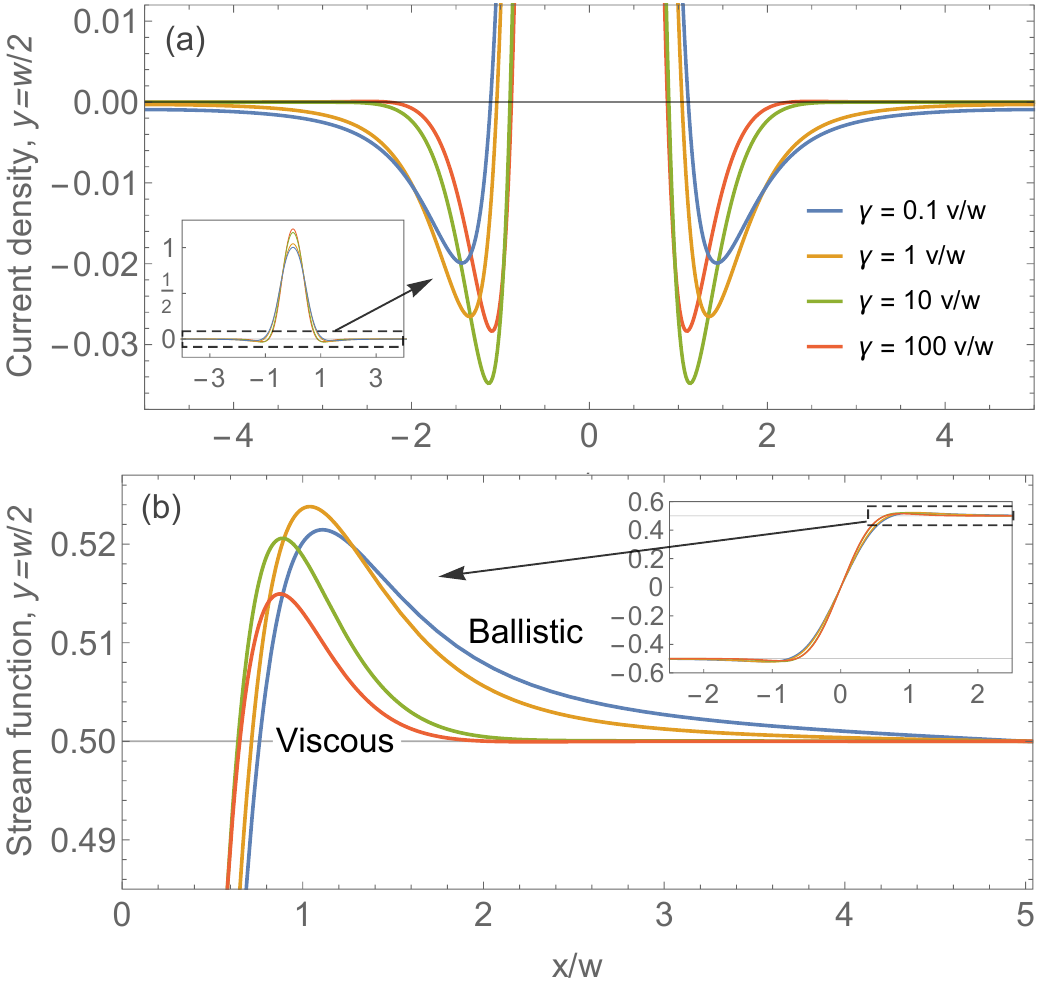} 
\caption{(a) Current backflow in the viscous, ballistic and intermediate regimes sampled on the line $y=w/2$ in the middle of the strip. The backflow magnitude is nearly the same in all regimes, being slightly larger in the intermediate regime compared to the viscous and ballistic regimes. Secondary vortices  in the viscous regime, manifested through multiple current sign reversals, are illustrated in the zoom-in shown in Fig.\ref{fig:Supp_Middle}. 
(b) Stream function $\phi_{y=w/2}(x)$ normalized to unit net current, $\phi_{y=w/2}(x)|^{L/2}_{-L/2}=1$ and a zoom-in detailing backflows for different regimes. The largest backflow magnitude occurs for $\gamma=1v/w$ (intermediate regime). 
}
\label{fig:NoDisorder}
\end{figure}

Here we consider electron flow in an infinite strip of width $w,-\infty<x<\infty$, $0<y<w$,
with a pair of slits on opposite sides serving as the injector and drain contacts, Fig.\ref{fig:streams}. 
For transport in a strip of width $w$ the relevant wavenumber is $k\approx 1/w$. Accordingly, in our simulation we use $\gamma=0.1v/w$ and $100 v/w$ to model the ballistic and hydrodynamic regimes; the values
$\gamma=1v/w$ and $10v/w$ are used to model the crossover between these regimes.

\begin{figure}
\centering{}\includegraphics[width=0.925\columnwidth]{{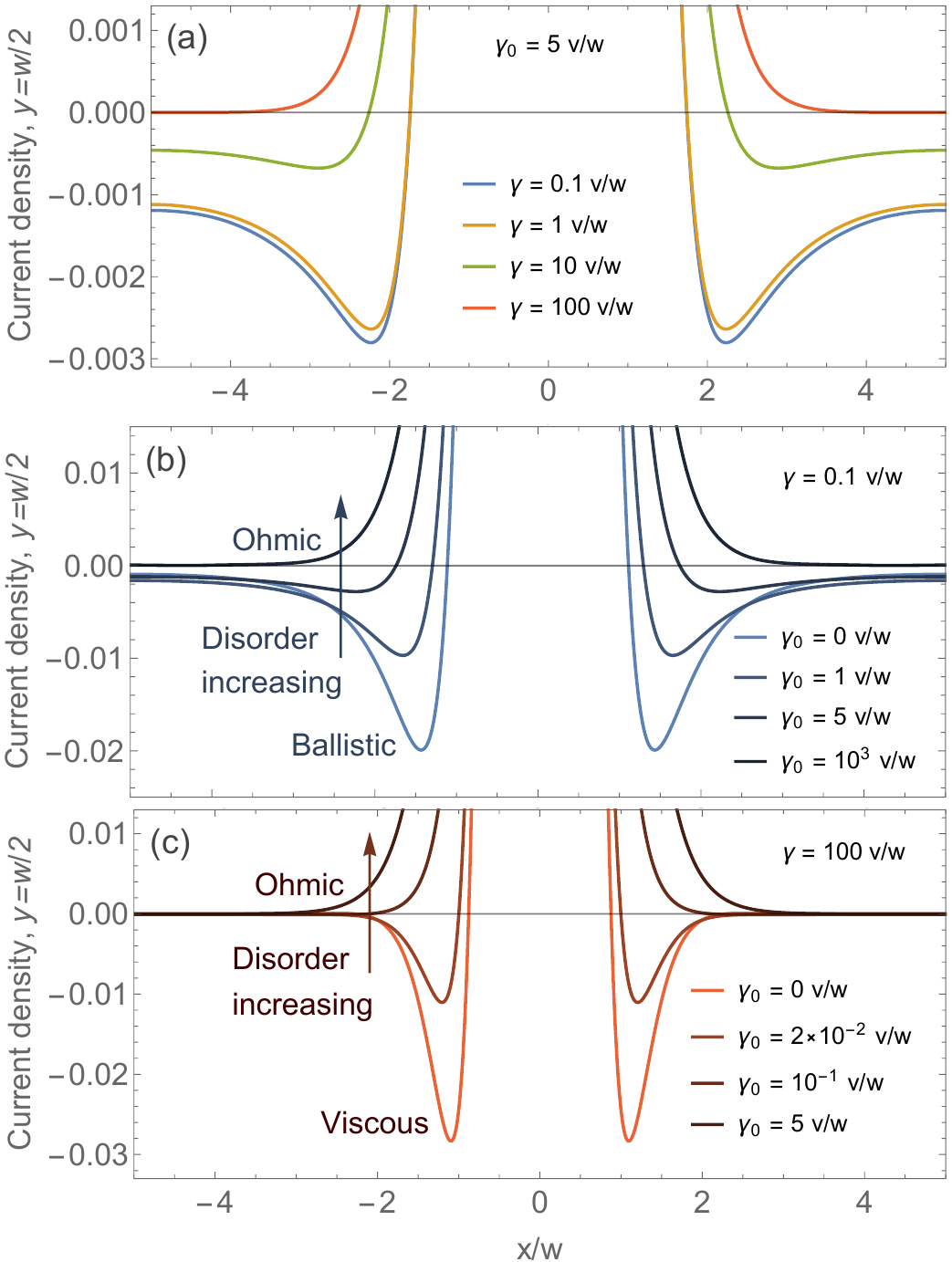}} 
\caption{\label{fig:WithDisorder} 
Suppression of current backflow and vorticity by ohmic dissipation in different regimes. 
(a) Current in the middle of the strip at a fixed disorder scattering rate $\gamma_{p}=5\,v/w$ and varying $\gamma$. 
The backflow and vorticity survive in the ballistic regime but are completely suppressed in the viscous regime. (b),(c) Weakening of the backflow upon increasing ohmic scattering in the ballistic and viscous regimes (realized at $\gamma=0.1\,v/w$ and $\gamma=100\,v/w$, respectively). Ballistic backflow is weakened roughly 2 times for $\gamma_{p}=1\,v/w$, and totally suppressed for $\gamma_{p}=10^3\,v/w$. Viscous backflow is considerably more fragile, being weakened roughly 2 times for $\gamma_{p}=2\times 10^{-2}\,v/w$ 
and completely suppressed for $\gamma_{p}=5\,v/w$.}
\end{figure}

We first consider the results for the disorder-free case [$\gamma_{p}=0$ in Eq.\eqref{sigma}]. After solving for currents at the strip boundary $y=0$ 
we use Eq.\ref{eq:j=sigma(E_ext-E_fic)} to find currents in the strip bulk. Vortices are revealed by current profile on a line in the middle of the strip $y=w/2$, pictured in Fig.\ref{fig:NoDisorder}(a). The key feature is the flow sign reversal:  
Outside the middle region where current flows from injector to drain 
there are regions where current flows against the applied field, signaling the presence of vortices. Interestingly, vortices are present 
in both the ballistic and viscous regime and 
have similar intensities with a little change at the crossover. 

One difference between vortices in the two regimes is in their spatial extent: ballistic vortices are about $\sim 2$ times wider than viscous vortices. Another (minor) difference is that current undergoes multiple  
sign reversals, indicating the presence of several vortices of opposite orientation (so-called Moffatt vortices~\cite{Moffatt1964,Semenyakin2018}). 
This confirms the presence of multiple vortices in the viscous regime 
pictured in Fig.\ref{fig:streams}. However, Fig.\ref{fig:NoDisorder} also indicates that the secondary vortices are extremely weak, 
illustrating that Fig.\ref{fig:streams} predicts correctly the flow geometry but misrepresents the magnitude of vorticity.  

To gain more insight, we consider the stream
function defined through $\vec{u}=\nabla\times(\phi(\vec r)\hat{z})=(\partial_{y}\phi(\vec r),-\partial_{x}\phi(\vec r))^{\rm T}$,
where $\vec{j}=en\vec{u}$ [see Fig.
\ref{fig:NoDisorder}(b)]. This quantity has a number
of useful properties. In particular, it quantifies the net integrated
backflow regardless of how far from the slit the backflow occurs
and the details of its spatial distribution 
and, as such, provides a meaningful comparison between different regimes. 
This quantity, shown in Fig.
\ref{fig:NoDisorder}(b) on the line in the middle
of the strip, $\phi_{w/2}(x)=\int_{0}^{x}j_{y}(\xi,w/2)d\xi$ 
indicates a larger swing for the ballistic flow (blue curve) 
than the viscous flow (red curve),
i.e. the backflow 
is actually somewhat stronger in the ballistic case than in the viscous
case. Yet, in the absence of disorder scattering, the predicted differences between ballistic and viscous vortices are probably not strong enough to unambiguously differentiate these regimes experimentally.

To the contrary, the ballistic and viscous vortices behave very differently in the presence of disorder scattering (ohmic dissipation). 
Specifically, as illustrated in Fig. \ref{fig:WithDisorder}, relatively weak disorder scattering proves sufficient to suppress viscous vortices, while having a relatively small impact on ballistic vortices, as discussed above. 
Electron-electron scattering is strongly temperature-dependent, behaving as $\sim T^2$ in 3D Fermi liquids \cite{BaymPethick} (also see recent work on 2D Fermi liquids \cite{Ledwith2019,tomogrph,Kryhin2022,Kryhin2023b,DasSarma2022}). 
Momentum relaxing scattering features little temperature dependence when it is due to scattering by disorder, and a linear $T$ dependence for electron-phonon scattering. 
Consequently, the parameters $\gamma$ and $\gamma_{p}$ can be adjusted by varying temperature, which 
makes the flow patterns sensitive to temperature. To illustrate this effect, we set $\gamma_{p}=5\, v/w$ and vary $\gamma$ as shown in Fig. \ref{fig:WithDisorder}(a). Notably, this $\gamma_{p}$ value is sufficient to fully suppress viscous backflow (indicated by the red line), while reducing ballistic backflow (depicted by the blue line) by only about $5$ times. The property of ballistic vortices being more resilient than viscous vortices in the presence of ohmic dissipation suggests a 
simple diagnostic for discriminating between the two regimes in experiments.

It is also instructive to consider how ballistic and viscous vortices, which have approximately equal intensity in the absence of ohmic dissipation, are suppressed as the disorder scattering rate $\gamma_{p}$ increases, see Fig. \ref{fig:WithDisorder} (b) and (c). In both cases we observe a transition to the ohmic flow regime that shows no backflow. 
Yet, the characteristic values of $\gamma_{p}$ above which the flow becomes effectively ohmic are very different for the two cases. Vortices in ballistic regime are quite robust in the presence of ohmic dissipation
and can sustain disorder scattering as high as $\gamma_{p}=10^{3}\,v/w$. 
For example, dissipation as small as $\gamma_{p}=2\times10^{-2}\,v/w$ results in a loss of the second (Moffatt) vortex and weakens the backflow amplitude 2 times, while for the ballistic case a similar reduction of the backflow happens for $\gamma_{p}=1\,v/w$.
These values are in agreement with the simple estimates given above in Eq.\eqref{eq:conductivities_a_b}. 

Systems in which electron hydrodynamics is currently being investigated, such as graphene and GaAs 2D electron gases\cite{Kumar2017,Bandurin2016,Sulpizio2019,Ku2020,Braem2018,Vool2021,Zeldov2022,Palm2024}, feature very low disorder concentration. Consequently, electron-phonon scattering becomes the primary mechanism for momentum relaxation, described by the rate $\gamma_p$ scaling linearly with temperature, $\gamma_p\sim T$. The electron-electron (ee) scattering rates behave as $\gamma\sim T^2$ with a relatively large prefactor, giving ee collision mean free path which is shorter than the el-ph mean free path at low $T$.  As temperature grows, this creates a range of lengthscales at which viscous regime sets in. It is interesting to note that 2D electron systems support a family of long-lived excitations with long memory times that feature decay rate $\gamma\sim T^4$ \cite{Ledwith2019,tomogrph,Kryhin2022,Kryhin2023b,DasSarma2022}. The occurrence of such excitations makes electron hydrodynamics non-Newtonian, featuring scale-dependent viscosity \cite{Kryhin2023,Kryhin2023b}. This impacts the behavior at low temperatures and modifies the ballistic-to-viscous crossover, but, as we presently believe, has little impact on vortices.  Then, as temperature continues to rise and the el-ph mean free path becomes shorter, eventually the ohmic regime takes over. Therefore, we expect ballistic transport and strong vortices at low and intermediate temperatures. As temperature rises, vortices initially persist when entering the viscous regime, but at higher temperatures, they are suppressed by momentum-relaxing electron-phonon scattering.


\section{Boundary value problem for vortical flows in a strip geometry}
\label{sec:strip geometry}

Here we detail the procedure used to evaluate the nonlocal response in the strip geometry. This is done by solving the integral equation for nonlocal transport, Eq. \eqref{eq:j=sigma(E_ext-E_fic)}, derived 
by replacing boundary conditions with a fictitious electric field. In this case, 
Eq. \eqref{eq:j=sigma(E_ext-E_fic)} reads:
\begin{align} \label{eq:B_12}
j_{\alpha}(\vec r)= & j_{0}\delta_{\alpha,y}-\lambda\int\limits_{\mathcal{B}_1\cup\mathcal{B}_2}  dx^{\prime}\sigma_{\alpha\alpha'}\left(\vec r-\vec{x^{\prime}}\right)j_{\alpha'}(\vec{x^{\prime}})
,
\end{align}
where $\mathcal{B}_1$ and  $\mathcal{B}_2$
denote 
strip boundaries with the intervals of the slits excluded, 
\begin{align} 
& \mathcal{B}_{1,2}=\{|x^\prime|>w/2\}_{y^\prime=0,w}.
\end{align} 
%
The term $j_{0}\delta_{\alpha,y}$ in Eq.\eqref{eq:B_12} represents an externally applied current which is perpendicular to the strip (see Eq.\eqref{eq:j=sigma(E_ext-E_fic)} and discussion beneath it).

In our analysis, we employ two spatial mirror symmetries. The horizontal mirror positioned at the strip middle line $y=w/2$ 
maps the upper and lower edges of the strip and the slits onto one another. The vertical mirror positioned at $x=0$ maps each of the edges and the slits onto themselves. These symmetries impose relations between the current components within the strip:
\begin{align}
&j_{y}(y,x)=j_{y}(y,-x),\quad j_{y}(y,x)=j_{y}(w-y,x); \label{eq:mirror_symmetries_y}\\
&j_{x}(y,x)=-j_{x}(y,-x),\quad j_{x}(y,x)=-j_{x}(w-y,x). \label{eq:mirror_symmetries_x}
\end{align}
We will focus on the currents at the edges and use these symmetry relations to simplify the problem by expressing variables on one edge through those on the other edge.



We first consider the $y$ component of the current. Although Eq.\eqref{eq:B_12} can, in principle, be solved directly in position space, it is more convenient here to switch to a mixed representation. This approach uses Fourier components along the strip (the $x$ direction) while retaining the direct-space representation perpendicular to the strip (the $y$ direction). By replacing the nonlocal conductivity with its Fourier transform, as in Eq.\eqref{eq:sigma_alpha_beta}, the equation for the $y$ component of the current becomes:
\begin{align}
& j_{y}(x,y)  =j_{0}-\lambda \int dx'\chi (x') j_{y}(x',0)
\la e^{ik_{x}(x-x')}\sigma_{yy}(\vec k)\ra_{+}
\nonumber
\\
& -\lambda \int dx'\chi (x') j_{x}(x',0)
\la e^{ik_{x}(x-x')}\sigma_{yx}(\vec k)\ra_{-}
, %
\end{align}
where $\la...\ra_{\pm}$ is a shorthand for $\int\frac{dk_{x}dk_{y}}{(2\pi)^{2}}...F_\pm(k_y)$, with  $F_\pm(k_y)=e^{ik_{y}y}\left(1\pm e^{-ik_{y}w}\right)$. 
For conciseness, we have extended the integral over $x^\prime$ to the entire axis $-\infty<x'<\infty$, introducing a window function defined as $\chi\left(x^{\prime}\right)=1$ when $x^{\prime}\geq w/2$, and zero for $x^{\prime}< w/2$. Also, we have used spatial mirror symmetries of the current components, Eqs. \eqref{eq:mirror_symmetries_x}, \eqref{eq:mirror_symmetries_y}. 

Next, we carry out a Fourier transform over $x$, $j_{y}(q,y)=\int j_{y}(x,y)e^{-iqx}dx$, to obtain
\begin{align} 
 j_{y}(q,y)  &=2\pi j_{0}\delta(q)-\lambda\int dx^{\prime}\chi\left(x^{\prime}\right)j_{y}(x^{\prime},0)e^{-iqx^{\prime}}
\nonumber\\
& \times \int\frac{dk_{y}}{2\pi}\left(e^{ik_{y}y}+e^{ik_{y}\left(y-w\right)}\right)\sigma_{yy}(q,k_{y}) \nonumber \\
 & -\lambda \int dx^{\prime}\chi\left(x^{\prime}\right)j_{x}(x^{\prime},0)e^{-iqx^{\prime}}
\nonumber\\
&\times  \int\frac{dk_{y}}{2\pi}\left(e^{ik_{y}y}-e^{ik_{y}\left(y-w\right)}\right)\sigma_{yx}(q,k_{y}).
\nonumber
\end{align}
Here $\delta(q)$ is a 1D delta function, and 
we used that the quantities $\sigma_{\alpha\alpha'}(q,k_{y})$ are $k_{y}$-even for $\alpha=\alpha'$ and $k_{y}$-odd for $\alpha\ne\alpha'$.  
%
%
Because 
of this property, in the second term $e^{ik_{y}y}+e^{ik_{y}\left(y-w\right)}$ can be replaced 
with $\cos\left(k_{y}y\right)+\cos\left(k_{y}\left(w-y\right)\right)$, whereas in the third term $e^{ik_{y}y}-e^{ik_{y}\left(y-w\right)}$ can be replaced with $i\left(\sin\left(k_{y}y\right)+\sin\left(k_{y}\left(w-y\right)\right)\right)$, as the remaining parts vanish due to parity of the expressions under the integrals. Further, we use the fact that the integrals over $x^{\prime}$ can be expressed as a convolution in the $q$ space, such as
\be
\int dx^{\prime}\chi\left(x^{\prime}\right)j_{y}(x^{\prime},0)e^{-iqx^{\prime}}=\tilde{\chi}\left(q\right)\ast j_{y}(q, y = 0)
.
\ee
In a similar way equations for the $x$ component of current can be derived. Passing to the mixed representation as above, and introducing the quantity $j_{x}(q,y)=\int dx e^{-iqx} j_x(x,y)$, 
we  obtain a full set of equations:
%
\begin{align}
j_{\alpha}(q,y) = 2\pi 
j_{0}\delta(q) \delta_{\alpha, y}
 - \lambda \Sigma_{\alpha \alpha^\prime}(q)\lb\tilde{\chi}\left(q\right)\ast j_{\alpha^\prime}(q)_{y = 0}\rb,
 \label{j alpha (q)}
\end{align}
where $j_{\alpha^\prime}(q)_{y = 0}$ is current on the strip edge and the dependence on $y$ comes through the quantities
%
\begin{align}
&\Sigma_{yy}(q)=\int\frac{dk_{y}}{2\pi}
\lp e^{ik_{y}y}+e^{ik_{y}(w-y)}\rp
\frac{\sigma\left(\kappa\right)q^{2}}{\kappa^2};\nonumber\\
&
\Sigma_{yx}(q)=
-\int\frac{dk_{y}}{2\pi}\lp e^{ik_{y}y}-e^{ik_{y}(w-y)}\rp
\frac{\sigma\left(\kappa\right)qk_{y}}{\kappa^2} 
\label{eq:Sigma_y}\\
&\Sigma_{xx}(q)=\int\frac{dk_{y}}{2\pi}
\lp e^{ik_{y}y}-e^{ik_{y}(w-y)}\rp
\frac{\sigma\left(\kappa\right)k_{y}^{2}}{\kappa^2};\nonumber\\
&
\Sigma_{xy}(q)=
-\int\frac{dk_{y}}{2\pi}\lp e^{ik_{y}y}+e^{ik_{y}(w-y)}\rp
\frac{\sigma\left(\kappa\right)qk_{y}}{\kappa^2}
, 
\label{eq:Sigma_x}
\end{align}
%
where we introduced notation $\kappa=\sqrt{q^{2}+k_{y}^{2}}$. When using these relations to evaluate 
$\Sigma_{\alpha \alpha^\prime}(q)$, it is important to account for 
the parity of the quantities integrated over $k_y$. This parity is such that only the $k_y$-even and $k_y$-odd parts 
of $e^{ik_{y}y}\pm e^{ik_{y}(w-y)}$ 
contribute to $\Sigma_{xx}(q)$, $\Sigma_{yy}(q)$ and 
$\Sigma_{xy}(q)$, $\Sigma_{yx}(q)$, respectively. Accordingly, we can replace the quantities 
of $e^{ik_{y}y}\pm e^{ik_{y}(w-y)}$ with 
\be
\cos k_{y}y \pm \cos k_{y}(w-y)
\ee 
under the integrals giving $\Sigma_{xx}(q)$, $\Sigma_{yy}(q)$ and with 
\be 
i\sin k_{y}y \pm i\sin k_{y}(w-y)
\ee 
in the integrals giving $\Sigma_{xy}(q)$, $\Sigma_{yx}(q)$. The significance of this property is that some of the integrals that technically diverge are nulled by parity, and the sine and cosine versions of the expressions for $\Sigma_{\alpha \alpha^\prime}(q)$ know about it.


The relations in Eqs.\ref{j alpha (q)} 
provide `bulk/boundary relations' that link currents in the interior of the strip $0<y<w$ and currents at the lower edge $y=0$. To determine currents in the strip interior we first solve for currents at the edge, and then use the above bulk/boundary relations to find currents in the entire strip. This procedure is detailed and illustrated in Sec.\ref{sec:boundary}.

\begin{figure}[t]
\includegraphics[width=0.8\columnwidth]{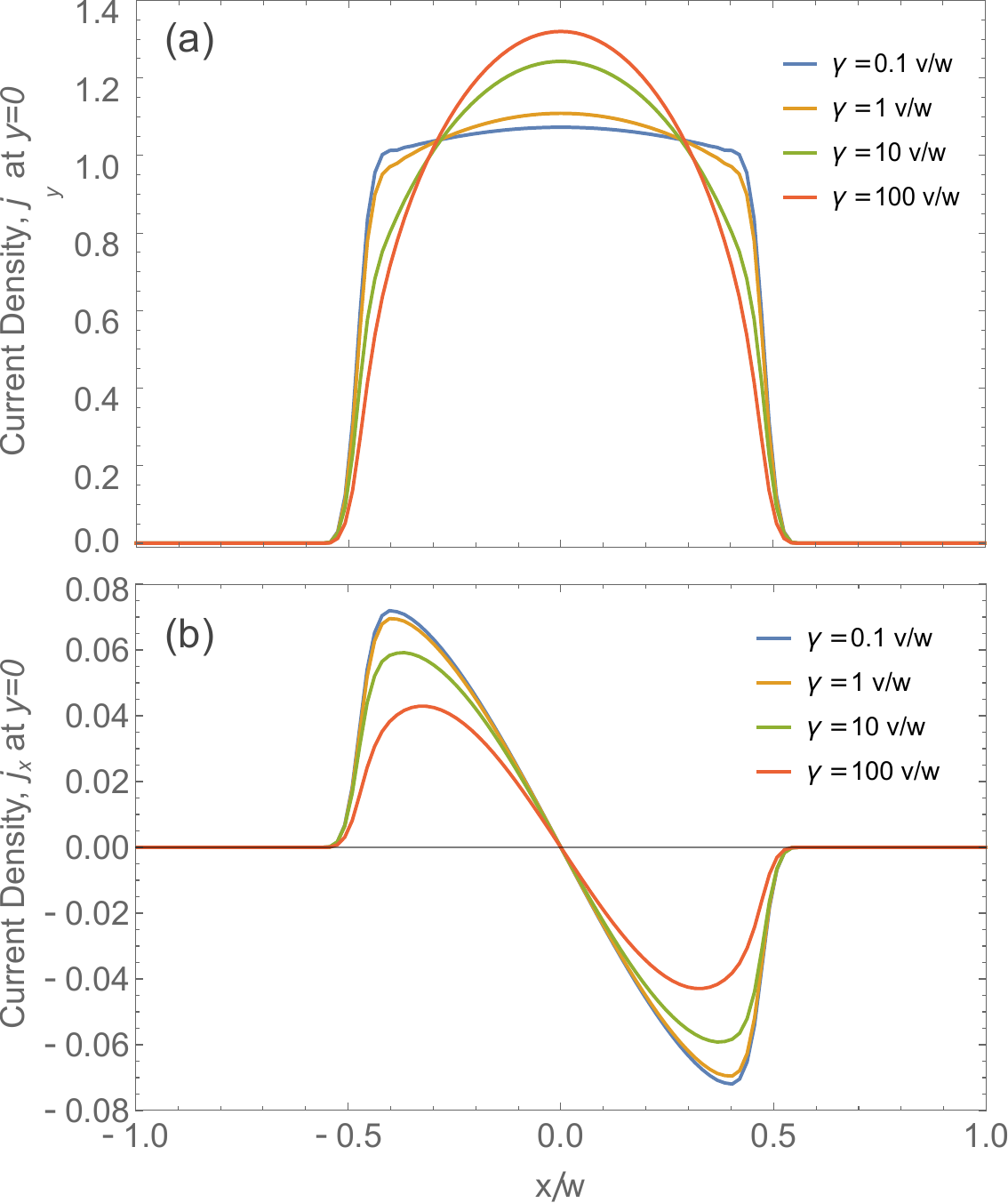}  
\caption{\label{fig:NoDisorder_y=0} Current distributions at the boundary $y=0$ in the dissipationless dynamics $\gamma_{p}=0$ for the viscous, ballistic and intermediate regimes. (a), (b) show the current components along the $y$ and $x$ axes, respectively 
} 
\end{figure}
\begin{figure}[t]
\includegraphics[width=0.8\columnwidth]{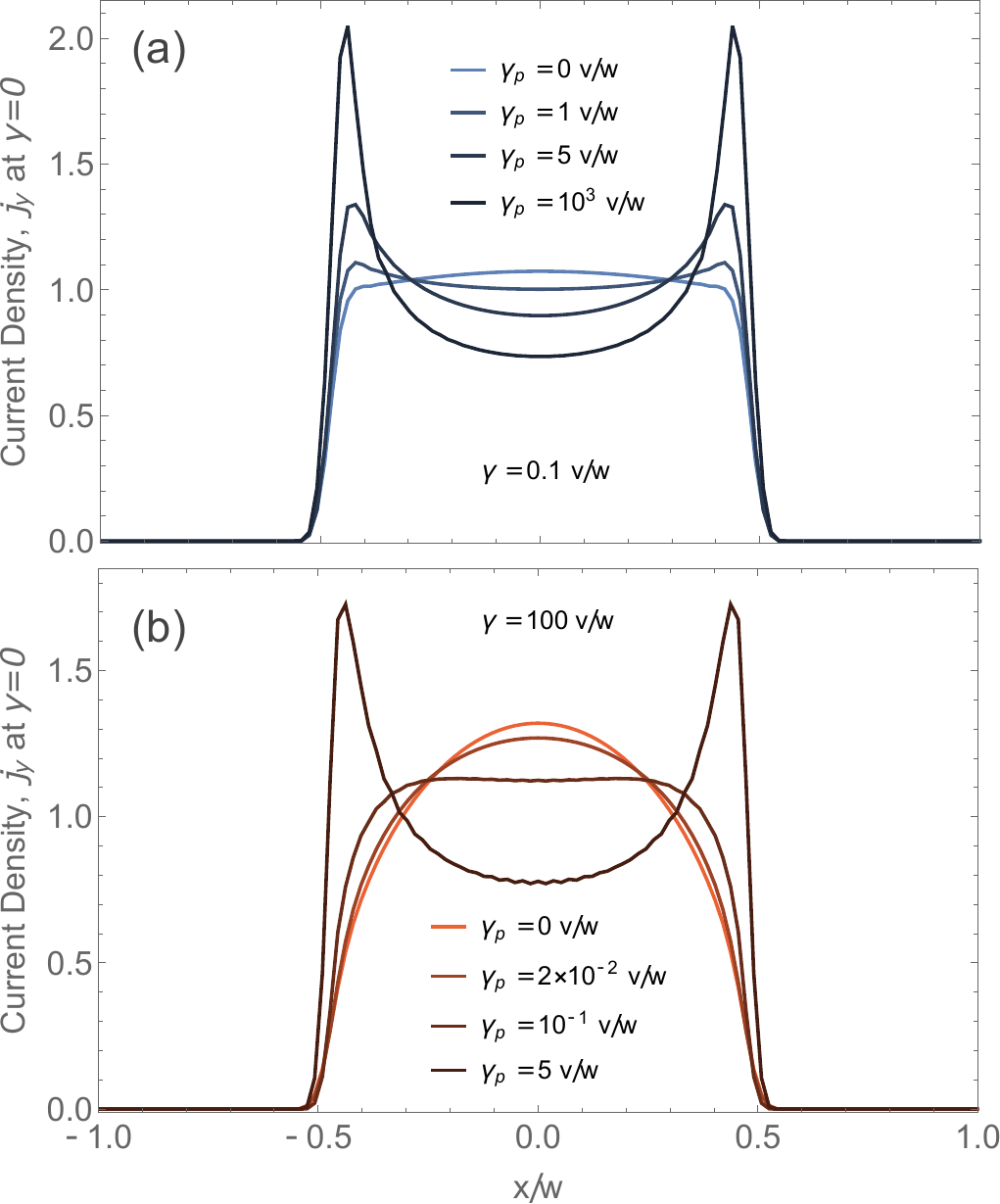} 
\caption{ \label{fig:WithDisorder_y=0} Current distributions along the $y$ axis at the boundary $y=0$ for increasing disorder scattering (a) ballistic to ohmic crossover (b) viscous to ohmic crossover
}
\end{figure}

\section{Finding currents at the strip boundary}
\label{sec:boundary}

Here we detail the approach used for solving Eq.\eqref{j alpha (q)}. The right-hand sides of these equations contain only the currents at the lower boundary $y=0$. Therefore, we set $y=0$ to obtain a pair of coupled linear integral equations for $j_x(q,0)$ and $j_y(q,0)$. For 
conciseness, in this section we will suppress the "0" label, denote current components simply as $j_{x,y}(q)$.
 
To solve these equations we need to address several technical issues. First,
in the absence of ohmic dissipation ($\gamma_{p}=0$) the conductivity $\sigma(k)$ has a divergence at $k=0$ [see Eq.\eqref{eq:conductivities_a_b}]. This translates intoa divergence in the mixed representation. In particular, the quantity $\Sigma_{yy}(q)$ diverges when $q\to0$. 
\begin{align}
\Sigma_{yy}\left|{}_{q\to 0}\right.\approx \int \frac{dk_y}{2\pi} \frac{8\gamma D}{v^2 (k_y^
2+q^2)} \frac{q^2}{k_y^2+q^2}= \frac{2D\gamma}{v^2} \frac{1}{|q|}
\end{align}
One way to eliminate this problem in numerical analysis is to introduce an
infinitesimal $\gamma_{p}$. Another way is to multiply the $y$ component of Eq.\eqref{j alpha (q)} by $\Sigma_{yy}^{-1}(q)$, which provides a regularization since $\Sigma_{yy}^{-1}(0)=0$ (see Ref. \cite{HGuo2017}). 
In order to treat all the regimes on equal footing here we adopt the latter approach.

However, as noted in Ref. \cite{HGuo2017}, where this approach has been successfully used to analyze transport in a constriction, there is a subtle point to be considered. 
Namely, 
since $\Sigma_{yy}^{-1}(0)=0$,
the term
\begin{align}
j_{0}\Sigma_{yy}^{-1}(q)\delta(q) 
\label{eq:nondiverging_term}
\end{align} 
appears to vanish at all $q$, both zero and non-zero, due to the properties of the $\delta$-function. Naively, this poses a conundrum, because the equations appear to loose the information about the injected current. To circumvent this issue, we make the system size $L$ finite, which transforms $2\pi\delta(q)$ into $2\sin(qL/2)/q$ as a Fourier transform of the external current which is constant within our system and nulls outside of it. Now, we are left with an expression of the form $|q|\cdot 2\sin(qL/2)/q$, which we want to transform back into a suitably normalized delta function. 
We can find the normalization constant by integrating this function over $q$:
\begin{align}
\label{eq:delta_function_normalization}
\int dq |q|\, 2\sin(qL/2)/q = 8/L,
\end{align}
where the integral $\int_0^{\infty} \sin (q L/2) dq$ was evaluated by introducing an  exponentially decaying factor $\exp(-\epsilon q)$ and then subsequently taking the limit $\epsilon\to 0$. Finally, after all manipulations, we can legitimately write:
\begin{align}
  2\pi j_{0}\Sigma_{yy}^{-1}(q)\delta(q)=  2\pi \mu \delta(q), \quad \mu = \frac{2}{\pi L}\frac{v^2}{D\gamma}j_0
  \end{align}  
  The quantity $\mu$ defined here corresponds to the parameter $\mu$ introduced in Ref. \cite{HGuo2017}, taking the same value. This derivation elucidates its natural emergence in this context and proves its finiteness.
We also note, that in the presence of ohmic losses, the problem with divergence in $\Sigma_{yy}(q)$
does not occur, and the equations can be treated without multiplying by 
$\Sigma_{yy}^{-1}(q)$. 



The second issue that needs to be addressed is that with $\Sigma_{xx}(q)$. This quantity, for $y=0$, diverges
logarithmically for arbitrary $q$. The divergence originates from
$k_{y}\to\infty$, i.e. the small length scales. From physical
perspective, a UV divergences can be treated by introducing a suitable regularization. 
In this case, a UV regularization is implemented by assuming that the boundary is a blurred-out line that has a finite thickness $a$. 
In this case, for a small enough $a$, we identify $j_{x}(q,a)\approx j_{x}(q,0)$,
which allows to evaluate $\Sigma_{xx}(q)$ by plugging into Eq.\eqref{eq:Sigma_x} $y=a$ instead of $y=0$: 
\begin{align}
\Sigma_{xx}(q)&=\int\frac{dk_{y}}{2\pi}\big[\cos\left(k_{y}y\right)-\cos\left(k_{y}\left(w-y\right)\right)\big]\big|_{y=a}
\nonumber\\
& \times \frac{\sigma\left(\kappa\right)k_{y}^{2}}{\kappa^2} 
,\quad \kappa=\sqrt{q^{2}+k_{y}^{2}}. \label{eq: sigma_xx with cutoff}
\end{align}
This integral converges for large $k_{y}$ and shows a logarithmic
dependence on the scale $a$.
 
The origin of this logarithmic dependence is the well-known effect of a log-divergent mean free path in clean metals, which occurs when carrier momentum relaxation is dominated by surface scattering\cite{Fuchs1938, Sondheimer1952}. This divergence arises from 'grazing electrons' traveling at small angles to the surface, allowing them to avoid surface collisions for extended periods. In metals, the upper cutoff for this divergence is determined by the Fermi momentum $k_F$, represented here by the quantity $a^{-1}$. The lower cutoff is due to large lengthscales set by the system length, represented here by the inverse wavenumber $k_{x}^{-1}$.


\begin{figure}[t]
\includegraphics[width=0.8\columnwidth]{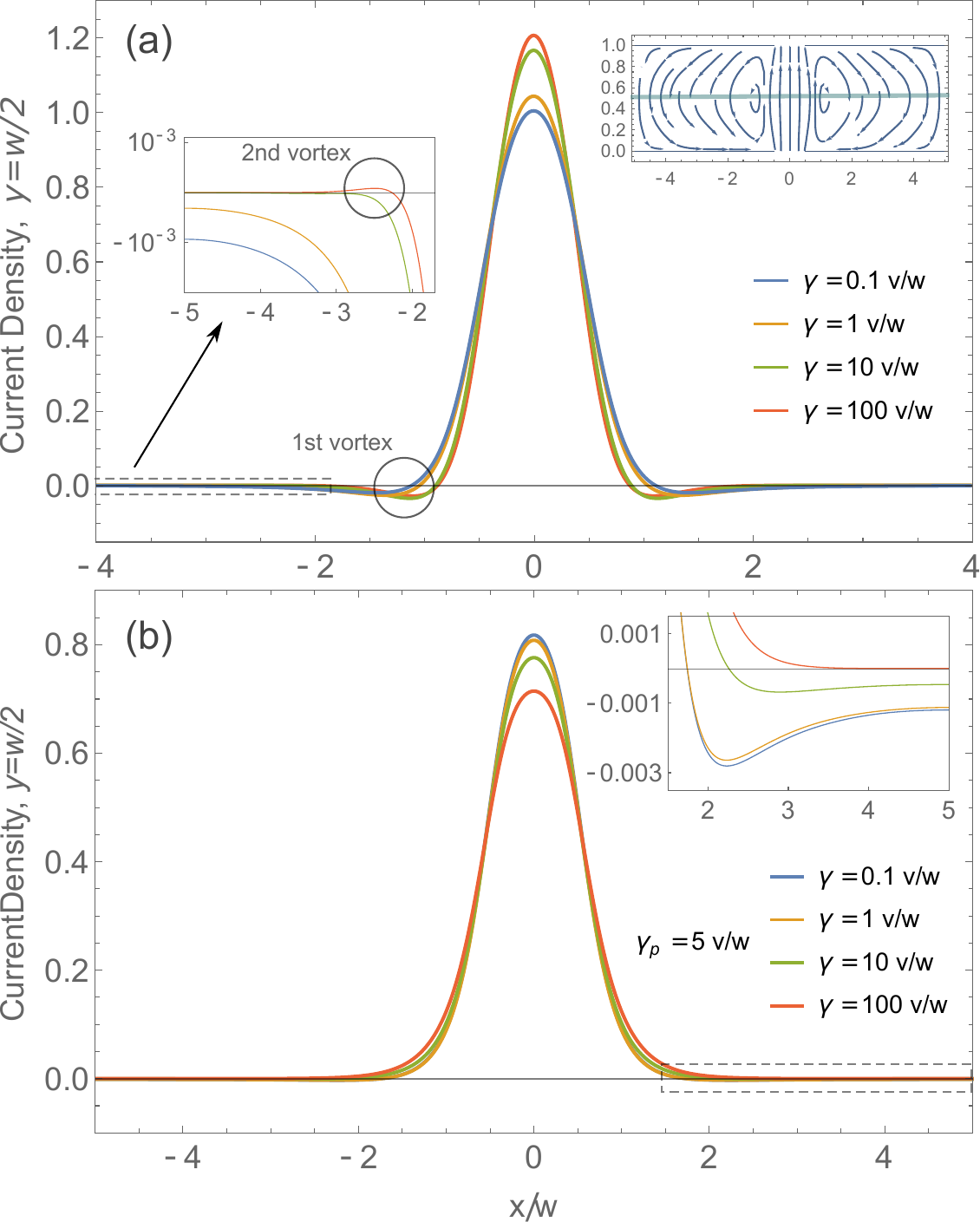} 
\caption{ (a) Current distributions on the strip middle line in the absence of ohmic dissipation. 
The line at which the distributions are calculated is shown in the right inset  superimposed with the flow in the strip.
Circles in the main panel and the left inset highlight sign reversals due to vortices. 
The lower circle highlights the primary vortex present for all the regimes, the upper circle in the left inset highlights the 2$^{\text{nd}}$ vortex of an extremely small amplitude present only in for $\gamma=100 v/w$, corresponding to an extreme viscous regime. 
(b) Same as in (a)  in the presence of finite ohmic dissipation  $\gamma_{p}=5 v/w$ the same for all $\gamma$ values. The inset in (b) is a zoom-in on the backflow region due to the primary vortex. While in the viscous regime vorticity  is completely suppressed at this level of ohmic dissipation, in the ballistic regime it is  reduced roughly $5$ times but remains finite. 
}
\label{fig:Supp_Middle}
\end{figure}

Continuing with the analysis, after making the adjustments described above, we proceed to solve the integral equations for current, Eq.\eqref{j alpha (q)}. Because of the convolution, these equations cannot be solved analytically, and a numerical approach is required. 
We therefore introduce an interval $-L/2<x<L/2$ on the $x$-axis, discretized with a 
spacing $\Delta x = L/N$ with a large enough $N$, here taken to be $N=571$. For the functions in this interval we assume periodic boundary conditions. In Fourier representation, these functions are sums of harmonics with discrete wavenumbers on the dual lattice 
\begin{align}
q_i= \left(i-\frac{N+1}{2}\right)\frac{2\pi}{L}
,
\quad i=1,2, ..., N,
\label{eq:mesh}
\end{align}
with a step size $2\pi/L$. In our numerical calculation, when treating Eq. \eqref{eq: sigma_xx with cutoff}, we 
take $a=\Delta x/10$. 

After this discretization, the currents $j_{x}(q)$, $j_{y}(q)$ become  $N$-component vectors, and the convolution becomes a linear operator representing the corresponding integral by a $N\times N$ matrix: $\tilde{\chi}\left(q\right)\ast j_{y}(q)=\hat{M}_{\chi}j_{y}(q)$. 
We solve our equations on the dual lattice, Eq. \eqref{eq:mesh}, approximating integrals as Riemann sums. An inverse Fourier transform is then carried out to find the currents in interval $-L/2<x<L/2$ in real space. This allows us to rewrite the current equations as:
%
\begin{align}
j_{y}(q) &=  2\pi j_0 \delta_{q,0} -\lambda\Sigma_{y \alpha^\prime}(q) \hat{M}_{\chi} j_{\alpha^\prime}(q)
\label{jy(q,0)}\\
j_{x}(q) & =-\lambda\Sigma_{x\alpha^\prime}(q)\hat{M}_{\chi}j_{\alpha^\prime}(q) 
\label{jx(q,0)}
.
\end{align}
%
From Eq.\eqref{jx(q,0)} we express 
\begin{align}
 & j_{x}(q)=-\lambda\left(1+\lambda\Sigma_{xx}(q) \hat{M}_{\chi}\right)^{-1}\Sigma_{xy}(q) \hat{M}_{\chi}j_{y}(q) 
 .
\label{jx(q,0)_fin}
\end{align}
Then we plug it into Eq.\eqref{jy(q,0)} multiplied by $\Sigma^{-1}_{yy}(q)$. This gives
\begin{align}
&j_{y}(q)=  2\pi \mu 
\left[\Sigma^{-1}_{yy}(q)+\lambda\hat{M}_{\chi}-\lambda^2\Sigma^{-1}_{yy}(q)\right.
\label{jy(q,0)_fin}\\ \nonumber
&\left. \times \Sigma_{yx}(q)\hat{M}_{\chi}\left(1+\lambda\Sigma_{xx}(q)\hat{M}_{\chi}\right)^{-1}\Sigma_{xy}(q) \hat{M}_{\chi}\right]^{-1}\delta_{q,0}
\end{align}
which we then use in Eq.\eqref{jx(q,0)_fin} to find the $x$ component.
For the final step we carry out an inverse Fourier transform
of the currents. 



%

The results of this calculation are presented in Fig.\ref{fig:NoDisorder_y=0} and Fig.\ref{fig:WithDisorder_y=0}. 
Current at the boundary vanishes outside the slit, as expected, and has a profile within the slit that reflects the 
miscroscopic scattering mechanisms. The dependence $j_y(x)$ is nearly flat in the ballistic regime, as expected from Sharvin's phase space argument \cite{Naidyuk2005,Sharvin1965}, and acquires a convex profile as the el-el collision rate grows, see Fig.\ref{fig:NoDisorder_y=0}(a). In this limit, current drops as $x$ approches the slit edges, as expected for a viscous flow with no-slip boundary conditions, resembling velocity profile for Poiseuile flow. 

The component $j_x(x)$ exhibits a sign-changing profile 
which indicates that within the slit on the $y=0$ boundary the current flows towards $x=0$ vertical axis, see Fig.\ref{fig:NoDisorder_y=0}(b). As can be seen in Fig.\ref{fig:streams}, 
at a slightly larger $y$, manifested through the flow lines spreading out. This behavior persists in the strip interior up to the middle line $y=w/2$; above this line the current flow is a mirror image of that below the line $y=w/2$, such that 
$j_x(x,y)=-j_x(x,w-y)$ and $j_y(x,y)=j_y(x,w-y)$. 

Further, the profile $j_y(x)$ within the slit undergoes a peculiar transformation when ohmic losses are introduced, developing a double-horn structure in both the ballistic and viscous regimes as the disorder scattering increases, see Fig. \ref{fig:WithDisorder_y=0}(a,b). This behavior reflects the familiar effect of current crowding near sharp corners expected for ohmic transport, and is in agreement with previous work \cite{Qi2021,Guo_thesis_2018}.

Current density at the strip boundary, found as described above, is then used to find current distributions within the strip using Eq.\eqref{j alpha (q)} with suitably chosen $y$. Fig. \ref{fig:Supp_Middle} shows the overall current distribution on the line $y=w/2$ for the ballistic, viscous and crossover regimes, and highlights vortices. 
This can be compared to Figs. \ref{fig:NoDisorder} (a) and (b) which detail the backflow effect for these current distributions.
The circle in the left inset of Fig. \ref{fig:Supp_Middle}(a) marks the secondary vortex for the viscous regime ($\gamma=100\,v/w$). As discussed in Sec.\ref{sec:general method}, the current in the viscous regime undergoes multiple sign reversals due to the presence of several vortices with opposite signs. In general, viscous flows feature trains of infinitely many higher-order vortices (Moffatt vortices) \cite{Moffatt1964,Semenyakin2018}. However, these vortices are extremely weak compared to the primary vortices, making them insignificant from a practical standpoint.

\section{Consistency checks}
\label{sec:consistency}
To verify the robustness of the conclusions obtained for the conductivity model $\sigma(k)$ used throughout the discussion above and given in Eq.\eqref{sigma}, we also considered another model, for which the rates $\gamma_m$ scale as $m^2$: 
\be
\gamma_{m>1}=\gamma m^2,
\quad 
\gamma_{m=1}=\gamma_{p}. 
\ee
For $\gamma_{p}=\gamma$, these rates define a collision operator describing transport governed by small-angle scattering wherein relaxation occurs by angular diffusion of carrier distribution along the Fermi surface, 
\be 
I f(\theta)=-\gamma \p_\theta^2 f(\theta),
\ee 
where $\theta$ is the angle on the Fermi surface. The case $\gamma_{p}\ne\gamma$ models the regime in which momentum relaxation rate $\gamma_{m=1}$ is not directly linked to the relaxation rates for higher angular harmonics of carrier distribution. Evaluating the quantity $R(k)$ in this case we have 
\be 
\sigma'(k)=\frac{D}{\gamma_{p}+{R'}(k) }
,\quad
{R'}(k)=\frac{z}{4\gamma+\frac{z}{9\gamma+\frac{z}{16\gamma+\ldots}}}
.\label{eq:diffus_model}
\ee
In the long-wavelength limit $kv\ll \gamma$ we can expand ${R'}(k)$ to leading order in $k^2$ to obtain
\be 
\sigma'(k)=\frac{D}{\gamma_{p}+\frac{k^2v^2}{16\gamma}+O(k^4) }
.\label{eq:diffus_model_small_k}
\ee
To study the flow we have used the approach described in earlier sections. 
We found that the flow obtained for the model given in Eq. \eqref{sigma} 
features vortices of comparable strength for $\gamma$ and $\gamma_{p}$ both large and small compared to $v/w$. This confirms that vorticity is a robust quantity taking similar values in the ballistic and viscous phase. 

\begin{figure}[tb]
\includegraphics[width=0.99\columnwidth]{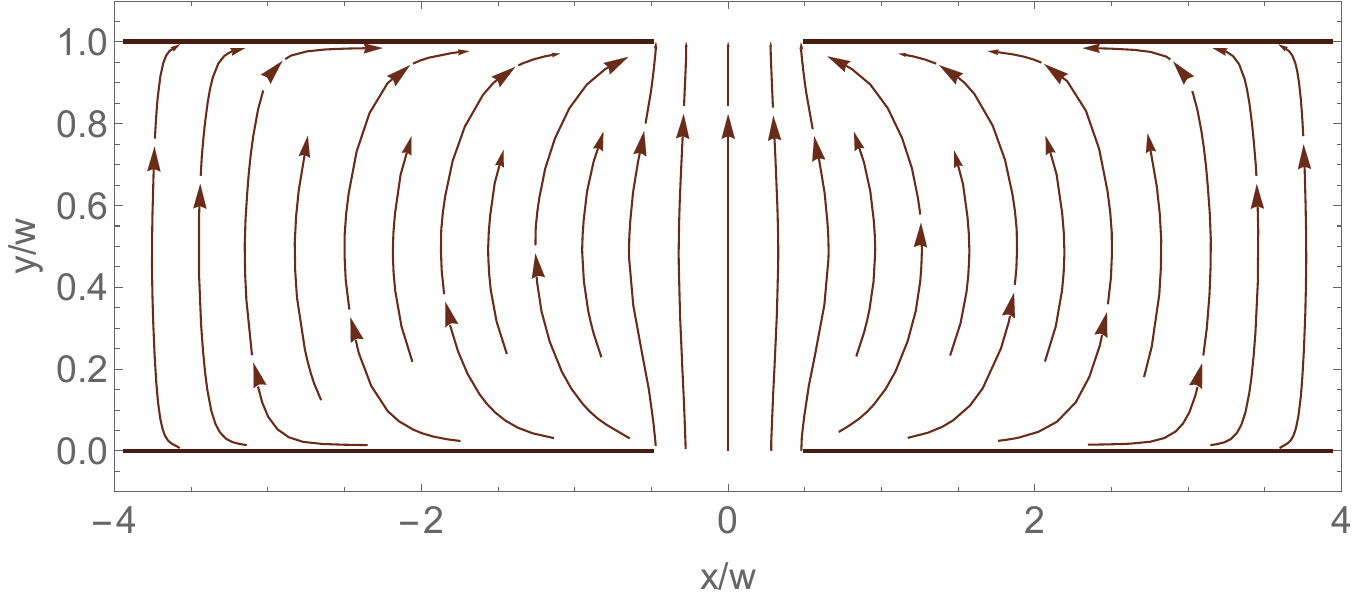}
\caption{\label{fig:ohmic} Current distribution in the ohmic regime obtained by the same procedure as the results for the ballistic and viscous regimes discussed above. In the ohmic regime, current flow is potential and thus features no vortices.}
\end{figure}	

The observed independence of flow patterns of the model $\sigma(k)$ specifics is not particularly surprising. 
Indeed, $\sigma(k)$ dispersion in the ballistic regime $\gamma\ll kv$ 
is identical for the models in Eqs. \eqref{sigma} and \eqref{eq:diffus_model}, matching 
the scaling law given in Eq.\ref{eq:conductivities_a_b} (a). Meanwhile, in the viscous regime, Eq.\eqref{sigma} can be approximated by the dependence given in Eq.\ref{eq:conductivities_a_b}(b), whereas Eq.\eqref{eq:diffus_model} can be approximated by Eq.\eqref{eq:diffus_model_small_k}. These expressions are of the same form, differing only in the choice of coefficients. In particular, 
in the absence of ohmic dissipation, $\gamma_{p}=0$, the conductivities differ only by a multiplicative constant factor. This explains why, upon changing the model, the macroscopic behavior of the flow remains the same in the ballistic and viscous limits. The differences between these models are present only in the intermediate regime, however, these are quite small. The only visible change is in the position of the ballistic/viscous phase boundary originating from the factor of sixteen in Eq.\eqref{eq:diffus_model_small_k}: 
The crossover in the  angular diffusion model, Eq.\eqref{eq:diffus_model}, takes place roughly four times faster with the rise of $\gamma$ than in the model used above, Eq.\eqref{sigma}; namely, the dynamics becomes effectively viscous for $\gamma\gtrsim 25$ and $\gamma\gtrsim 100$, respectively.

In addition, as another consistency check, we consider the extreme ohmic regime 
$\gamma_{p}\gg\gamma_{m\ge 1}$. This provides a useful comparison with the ballistic and viscous regimes discussed above. We can use a constant conductivity $\sigma(k)=D/\gamma_{p}$
to evaluate the integrals $\Sigma_{\alpha\alpha'}$ except for $\Sigma_{xx}$,
for which we still need to include the $k$-dependence in order to
control the convergence of the integrals. This leads to the current distribution
shown in Fig. \ref{fig:ohmic}. As expected, in the ohmic regime the flow is potential and thus vortex-free.


\section{Conclusions}

This article aims to develop a broad framework linking macroscopic vorticity in electron systems with microscopic interactions and scattering mechanisms. This is achieved by employing a wavenumber-dependent conductivity $\sigma(k)$ that accounts for realistic microscopic scattering processes. As an application of this approach, we clarify the relationship between nonlocal response and vortices across ballistic and hydrodynamic phases, 
illustrating it with vortical flows in a strip geometry. The qualitative behavior---such as similar vorticity intensity in ballistic and viscous regimes, the resilience of ballistic vortices to ohmic dissipation, and the comparatively fragile nature of viscous vortices---is expected to hold in any realistic geometry. Recent reports of vortices observed outside the viscous regime \cite{Zeldov2022,Palm2024} confirm the prediction of electronic vortices in the ballistic regime. The distinct dependence of vortical flows on electron-electron scattering versus ohmic dissipation serves as a clear indicator to differentiate the origins of vorticity in electron fluids. These findings establish the general relationship between vortices and nonlocal current-field response, highlighting the universal character of vortical flows in electron fluids.

We thank Margarita Davydova, Gregory Falkovich, Haoyu Guo and Eli Zeldov for helpful discussions. 
This work was supported by the Science and Technology Center for
Integrated Quantum Materials, NSF Grant No. DMR1231319; Army Research Office Grant W911NF-18-1-0116; and Bose Foundation Research fellowship.

\end{document}